\documentclass[12pt]{article}
\pdfoutput=1
\usepackage{jheppub}
\usepackage{verbatim}
\usepackage{amsmath}
\usepackage{amssymb}
\usepackage{amsthm}
\usepackage{slashed}
\usepackage{amsfonts}
\usepackage[utf8]{inputenc}
\usepackage[normalem]{ulem}
\usepackage{tikz}
\usetikzlibrary{patterns}
\usepackage{pgfplots}
 \usepgfplotslibrary{fillbetween}

\def\be{\begin{equation}}
\def\ee{\end{equation}}
\def\ba#1\ea{\begin{align}#1\end{align}}
\def\bg#1\eg{\begin{gather}#1\end{gather}}
\def\bm#1\em{\begin{multline}#1\end{multline}}
\def\bmd#1\emd{\begin{multlined}#1\end{multlined}}

\def\r{\rho}

\def\x{\xi}

\def\({\left(}
\def\){\right)}
\def\[{\left[}
\def\]{\right]}
\def\<{\langle}
\def\>{\rangle}

\newcommand{\bfig}{\begin{figure}\begin{center}}
\newcommand{\efig}{\end{center}\end{figure}}
\newcommand{\bi}{\begin{itemize}}
\newcommand{\ei}{\end{itemize}}

\newcommand{\Tr}{\mathrm{Tr}}

\theoremstyle{definition}

\DeclareMathOperator\erf{erf}

\begin{document}

%\subheader{empty}
\title{Probing phase transitions of holographic entanglement entropy with fixed area states}
\author[]{Donald Marolf,}
\author[]{Shannon Wang}
\author[]{and Zhencheng Wang}
\affiliation[]{Department of Physics, University of California, Santa Barbara, CA 93106, USA}
\emailAdd{marolf@ucsb.edu}
\emailAdd{shannonwang@ucsb.edu}
\emailAdd{zhencheng@ucsb.edu}

\abstract{ Recent results suggest that new corrections to holographic entanglement entropy should arise near phase transitions of the associated  Ryu-Takayanagi (RT) surface.  We study such corrections by decomposing the bulk state into fixed-area states and conjecturing that a certain `diagonal approximation' will hold.  In terms of the bulk Newton constant $G$, this yields a correction of order $O(G^{-1/2})$ near such transitions, which is in particular larger than generic corrections from the entanglement of bulk quantum fields.  However, the correction becomes exponentially suppressed away from the transition. The net effect is to make the entanglement a smooth function of all parameters, turning the RT `phase transition' into a crossover already at this level of analysis.

We illustrate this effect with explicit calculations (again assuming our diagonal approximation) for boundary regions given by a pair of disconnected intervals on the boundary of the AdS$_3$ vacuum and for a single interval on the boundary of the BTZ black hole.  In a natural large-volume limit where our diagonal approximation clearly holds, this second example verifies that our results agree with general predictions made by Murthy and Srednicki in the context of chaotic many-body systems.  As a further check on our conjectured diagonal approximation, we show that it also reproduces the $O(G^{-1/2})$ correction found Penington et al for an analogous quantum RT transition.  Our explicit computations also illustrate the cutoff-dependence of fluctuations in RT-areas.}

\maketitle

\section{Introduction}

The Ryu-Takayanagi (RT) \cite{Ryu:2006bv,Ryu:2006ef} prescription,  or more generally that of Hubeny-Rangamani-Takayanagi (HRT) \cite{Hubeny:2007xt}, computes the entanglement entropy in some region $R$ of a holographic CFT at leading order in the dual bulk Newton constant $G$.  To this order, the entropy is given by $A/4G$ in terms of the area $A$ of an extremal surface homologous to $R$ \cite{Lewkowycz:2013nqa}.  In addition, a  well-known correction at order $G^0$ is given by the entanglement of bulk fields \cite{Faulkner:2013ana}.

However, in the context of chaotic many-body systems it was recently noted that entanglement entropy can have extra correction terms near entanglement phase transitions \cite{Vidmar:2017pak,Murthy:2019qvb}.  In particular, motivated by \cite{Vidmar:2017pak}, Murthy and Srednicki studied energy eigenstates in systems satisfying the eigenstate thermalization hypothesis (ETH)  \cite{Murthy:2019qvb}.  Dividing the system into two spatial regions of volume $V_1$ and $V_2$ then yields a nontrivial entanglement entropy $S_{ent}(E)$.  Taking a large-volume limit and ignoring terms that scale no faster than the area of the interface between $V_1$ and $V_2$ allows one to define a corresponding  partition of the total energy, $E = E_1 + E_2$, between the two regions.  In this context,  for generic $V_1,V_2$, they
show the entanglement entropy $S_{ent}(E)$ to be approximated to exponential accuracy by the lesser of the microcanonical entropies $S_1(E_1),S_2(E_2)$ determined by the associated partition of the total energy $E = E_1 + E_2$ between the two regions.  But there is a larger correction of order $\sqrt{S_1} = \sqrt{S_2}$ near the transition where $S_1(E_1) =S_2(E_2)$.  Furthermore, the net effect of this correction is to make the entanglement a smooth function of all parameters, so that the apparent `phase transition' in fact becomes a crossover already at this level of analysis\footnote{In the strict limit of large volume the crossover occurs very quickly and one recovers the expected sharp phase transition.}.

Closely related physical settings have been considered in the holographic context for some time.  For example, one may consider a pure-state black hole, divide the boundary into regions $V_1,V_2$, and compute the HRT entropy; see e.g. \cite{Asplund:2014coa,Bao:2017guc}.  One then finds that the leading-order bulk RT/HRT computation describes a sharp RT/HRT phase transition with no analogue of the corrections described in \cite{Vidmar:2017pak,Murthy:2019qvb}.  This should not be a surprise as RT/HRT entropy is of order $1/G$ so the above $\sqrt{S}$ correction is only of order $G^{-1/2}$.  But such a correction should appear in a more complete study, and one might expect similar $O(G^{-1/2})$ corrections to arise near more general RT/HRT transitions as well.  These corrections are too large to arise from the entropy of bulk fields, and so must arise from some other aspect of the semiclassical approximation in the bulk.  A related $O(G^{-1/2})$ correction was recently discussed in \cite{Penington:2019kki} for an analogous quantum RT transition.

Our goal below is to provide a general description of such corrections near RT/HRT phase transitions using properties of the bulk fixed-area states introduced in \cite{Dong:2018seb} (see also \cite{Akers:2018fow}).
For simplicity, we focus on the time-symmetric (RT) case below where one may use real Euclidean path integrals.  However, we expect that the essential argument can be generalized to the more general HRT context using the Schwinger-Keldysh techniques of \cite{Dong:2016hjy}.
In particular, we decompose a general bulk into states in which we have simultaneously fixed the areas of all extremal surfaces satisfying the homology constraint (i.e., we have fixed the areas of all candidate RT surfaces).  For simplicity, we assume below that there are precisely two such extremal surfaces in a given such fixed-area state, and that their areas have been fixed to $A_1$ and $A_2$.   We then argue that the entanglement $S(A_1,A_2)$ in the associated fixed-area state $|A_1,A_2\rangle$ is given by RT up to corrections of order $G^0$, so that
\begin{equation}
S(A_1,A_2) = \frac{1}{4G}\min (A_1,A_2) + O(G^0).
\end{equation}
We also conjecture that -- again up to corrections of order $G^0$ -- the entanglement in a more general holographic state $|\psi \rangle = \int dA_1 dA_2 \psi(A_1,A_2) |A_1,A_2\rangle$ can be computed using a certain `diagonal approximation.'  When this conjecture holds, we show to leading order in $G$ that the von Neumann entropy is just the expectation value of $S(A_1,A_2)$ in the natural ensemble defined by the (normalized) state $|\psi\rangle$; i.e.
\begin{equation}
\label{eq:Sexp}
S = \int dA_1 dA_2 |\psi(A_1,A_2)|^2 S(A_1,A_2) + O(G^0).
\end{equation}
Evaluating this expression then gives the desired contribution at order $G^{-1/2}$, and with properties directly analogous to the correction of \cite{Murthy:2019qvb}.  Finally, we provide some evidence in support of our diagonal approximation by demonstrating agreement with the results of both \cite{Murthy:2019qvb} and \cite{Penington:2019kki}.

We begin in section \ref{Sec:Review} with a brief review of fixed area states.
General arguments for \eqref{eq:Sexp} and a statement of our diagonal-approximation conjecture are then given in section \ref{Sec:Correction}. The rest of the paper is devoted to more detailed computations of the effect, and to showing that our diagonal approximation reproduces results from \cite{Murthy:2019qvb} and \cite{Penington:2019kki}.  Section \ref{Sec:SingleInterval} consists of a warm-up exercise in which we study fixed-area states associated with a single interval in vacuum AdS$_3$.  While there is no phase transition in this context, results from this simple context will be useful studying examples of the above phase transition in section \ref{Sec:Examples}. The first example concerns a pair of intervals on the boundary of vacuum AdS$_3$, while the second involves a single interval on the boundary of the Ba\~nados-Teitelboim-Zanelli (BTZ) black hole \cite{Banados:1992wn,Banados:1992gq}. After taking a natural large volume limit, the latter context allows us to demonstrate explicit agreement between our BTZ results and the predictions of \cite{Murthy:2019qvb}.
A final part of section \ref{Sec:Examples} shows that we can also reproduce the $O(1/\sqrt{G})$ correction found in \cite{Penington:2019kki} for an analogous quantum RT transition.  We close with some final comments in section \ref{Sec:Disc}, and in particular discuss the cutoff dependence of fluctuations in RT-areas.

Closely related work has been done independently by Xi Dong and Huajia Wang \cite{Dong:2020iod}.  We have arranged with them to coordinate simultaneous postings of the original versions of the papers to the arxiv.

\section{Review of fixed area states}
\label{Sec:Review}
We now briefly review some basic properties of fixed area-states following \cite{Dong:2018seb}. In particular, after defining the fixed-area states, we will review their connection with the probability distribution $P(A_*)$ for a holographic state to have RT-area $A_*$, features of the semiclassical approximation for such states, and the simple form of their Renyi entropies.  All of these features will play important roles in the analysis of section \ref{Sec:Correction}.

We consider a CFT state $|\psi\rangle$ prepared by a Euclidean path integral over a manifold $M_{CFT}$ with boundary $\partial M_{CFT}$.  It is thus natural to think of $|\psi\rangle$ as a state on the surface $\partial M_{CFT}$.

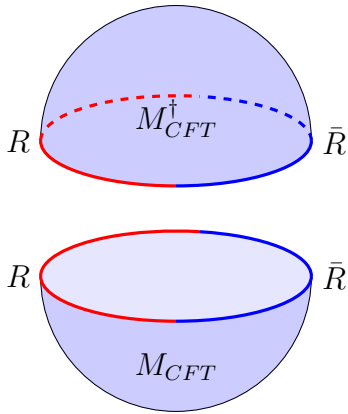
\begin{figure}
\centering
\begin{tikzpicture}[scale=0.6]
\draw [name path=A](-3,0) arc (180:360:3);
\draw [name path=B](-3,0) arc (180:360:3 and 1);
\draw[name path=C](-3,0) arc (180:0:3 and 1);
\tikzfillbetween[of=A and B]{blue, opacity=0.2};
\tikzfillbetween[of=C and B]{blue, opacity=0.1};
\draw [very thick,red] (-3,0) arc (180:270:3 and 1);\draw [very thick,red] (-3,0) arc (180:80:3 and 1);
\draw [very thick,blue] (3,0) arc (0:80:3 and 1);\draw [very thick,blue] (3,0) arc (0:-90:3 and 1);

\draw [name path=D](-3,3) arc (180:0:3);
\path [name path=E](-3,3) arc (180:0:3 and 1);
\path[name path=F](-3,3) arc (180:360:3 and 1);
\tikzfillbetween[of=D and F]{blue, opacity=0.2};
\draw [very thick,red] (-3,3) arc (180:270:3 and 1);\draw [dashed,very thick,red] (-3,3) arc (180:80:3 and 1);
\draw [dashed,very thick,blue] (3,3) arc (0:80:3 and 1);\draw [very thick,blue] (3,3) arc (0:-90:3 and 1);

\node at (0,-2) {$M_{CFT}$};\node at (0,3.5) {$M^{\dagger}_{CFT}$};
\node at (-3.5,0) {$R$};\node at (3.5,0) {$\bar{R}$};
\node at (-3.5,3) {$R$};\node at (3.5,3) {$\bar{R}$};
\end{tikzpicture}
\caption{The manifold $M_{CFT}$ (bottom) that we use in the Euclidean path integral to prepare our holographic state $|\psi\rangle$ and the CPT-conjugate manifold $M^\dagger_{CFT}$ (top). Sewing the two together along their boundaries defines the manifold $M_{double} : = M_{CFT}^\dagger M_{CFT}$.  If the state is time-symmetric, then $M^\dagger_{CFT}$ is equivalent to $M_{CFT}$, the two manifolds are exchanged by the relevant notion of time-reversal, and this symmetry leaves invariant the boundary $\partial M_{CFT} = \partial M^\dagger_{CFT}$ of $M_{CFT}, \partial M^\dagger_{CFT}$.  The surface $\partial M_{CFT}$ is partitioned into regions $R$ (red) and $\bar R$ (blue).}
\label{fig:MCFT}
\end{figure}

We suppose that $\partial M_{CFT}$ is partitioned into regions $R$ and $\bar{R}$.  For simplicity, we take the state to be invariant under a time-reflection symmetry that leaves fixed the surface $\partial M_{CFT}$.   Under the AdS/CFT correspondence, we may identify $M_{CFT}$ with the boundary of a bulk system, and we may similarly identify $\partial M_{CFT}$, $R$, $\bar{R}$ with corresponding (partial) surfaces in that boundary.  We will use $\partial R$ to denote the boundary between $R$ and $\bar{R}$ within $\partial M_{CFT}$.  The correspondence also tells us that the norm $\langle \psi | \psi \rangle$ can be computed using a Euclidean bulk path integral with boundary conditions defined by the closed manifold $M_{double} : = M_{CFT}^\dagger M_{CFT}$ defined by sewing together $M_{CFT}$ and its CPT-conjugate $M_{CFT}^\dagger$ along the common boundary $\partial M_{CFT}$; see figure \ref{fig:MCFT}.  The assumption of time-symmetry requires $M_{CFT}^{\dagger}$ to be equivalent to $M_{CFT}$, so that $\partial M_{CFT}$ is a surface of time-symmetry in $M_{double}$.

Roughly speaking, given a state $|\psi\rangle$ defined as above, we wish to define associated states $|\psi\rangle_{A_*}$ of fixed RT area by restricting the domain of integration to metrics for which the area $A_{\gamma_R}$ of the RT surface\footnote{A better approach which avoids the need to define an RT surface for off-shell metrics may be to build a path integral using the fixed-area action of \cite{Dong:2019piw}.  This action singles out a preferred surface whose area is to be fixed and then finds that the equations of motion require it to be an RT surface modulo imposition of the homology constraint.} $\gamma_R$ takes on a  definite value $A_*$, and by thus projecting  $|\psi\rangle$ onto the subspace with area $A_*$.
In this sense, the norm of a fixed area state is calculated by the path integral
\begin{equation}
\label{eq:fixednorm}
\begin{aligned}
[]_{A_*}\langle\psi|\psi\rangle_{A_*} &=\int \mathcal{D}g|_{A_{\gamma_R}=A_*} e^{-I[g]}\\
&=\int \mathcal{D} g d\mu e^{-I[g]-i\mu(A_{\gamma_R}[g]-A_*)}.
\end{aligned}
 \end{equation}
In the second line we have introduced a Lagrange multiplier $\mu$ to enforce the constraint on the area of $\gamma_R$. In practice, we will wish to restrict $A_{\gamma_R}$ to some window around $A_*$ where the width of the window is small compared to other scales of interest, but where the window still contains many area-eigenvalues. As a result, one should think of the measure $d\mu$ as being a broad Gaussian measure instead of being precisely flat.  However, we will take this measure to be sufficiently flat that its Gaussian nature can be ignored in the saddle-point approximation used below.

Due to our projection, the path integral \eqref{eq:fixednorm} is closely associated with the probability $P({A_*})$ for the holographic state $|\psi\rangle$ to have an RT area in the above window about $A_*$.  In particular, we have
\begin{equation}
P(A_*) = \frac{
{}_{A_*}\langle\psi|\psi\rangle_{A_*} }{
\langle\psi|\psi\rangle }.
\end{equation}

Since we will study \eqref{eq:fixednorm} in the saddle-point approximation, our task will be to find on-shell solutions to the Euclidean equations of motion. As is well known\footnote{Though see appendix A of \cite{Dong:2019piw} for a more complete justification.}, at this level the integral over $\mu$ and the term $-i\mu A_{\gamma_R}[g]$ in the exponent allow the insertion of an arbitrary conical defect (aka `cosmic brane') at the location of the RT surface.   The defect angle is to be chosen so that the saddle-point geometry $g_*$ satisfies the constraint $A_{\gamma_R} = A_*$. In the stationary phase approximation we thus find
\begin{equation}
\label{eq:fixednormstat}
{}_{A_*}\langle\psi|\psi\rangle_{A_*} \approx  e^{-I[g_*]}.
 \end{equation}
Note that $I[g_*]$ is the full gravitational action for $g_*$ and in particular includes a contribution from the delta-function curvature scalar on the conical singularity.

A priori, the form of \eqref{eq:fixednorm} suggests an imaginary conical defect angle $i\mu$, but as always the relevant saddles may not lie on the original contour of integration.  As a result, real defect angles (with imaginary values of our $\mu$) are allowed, and may arise with either sign.  Note that real $\mu_E = i\mu$ is in fact generally required for the stationary point $g_*$ to satisfy real Euclidean boundary conditions.  Thus $g_*$ is typically a real Euclidean metric, though it may contain either a conical deficit or a conical excess.  As discussed in \cite{Dong:2018seb,Dong:2019piw}, the location of the conical deficit should be thought of as the RT surface in the conical geometry.  We will thus refer to it as such below.

Since the classical actions $I(A_*) = I(g_*)$ are proportional to $1/G$, in the semiclassical limit $G\rightarrow 0$ the distribution $P(A_*)$ becomes sharply peaked about the most likely value $\bar A$.  This mostly likely values can be found by maximizing $P(A_*)$, or equivalently by minimizing the on-shell action with respect to $A_*$.  But minimizing the action in this way imposes the remaining Einstein equations on $\gamma_R$, and thus forbids any cosmic brane sources.  As a result, the most likely value $\bar A$ is just the area of $\gamma_R$ in the dominant bulk saddle $g_0$ associated with the path integral that computes the norm $\langle \psi | \psi \rangle$ \cite{Dong:2018seb,Dong:2019piw} without any a priori specification of areas.

Finally, we turn to considerations of entropy. Let us consider  the normalized density matrix $\rho_{A_*}$ on $R$ defined by the CFT dual to the bulk fixed-area state $|\psi\rangle_{A_*}$.  This density matrix may be written in the form
\begin{equation}
\label{rhoA0}
\rho_{A_*} = \frac{{\rm Tr}_{\bar R} \left( |\psi\rangle_{A_*} {}_{A_*}\langle \psi| \right)}{ {}_{A_*}\langle \psi | \psi\rangle_{A_*} },
\end{equation}
where in \eqref{rhoA0} we have used $|\psi\rangle_{A_*}$ to also denote the CFT dual to the bulk fixed-area state $|\psi\rangle_{A_*}$.
In the above semiclassical approximation, the freedom to tune the conical defect angle to enforce the constraint makes it straightforward to compute Renyi entropies $S_n(A_*) = \frac{1}{1-n} \ln {\rm Tr}_R \rho^n_{A_*}$.  In particular, the associated saddles $g_n(A_*)$ are just $n$-sheeted branched covers of the saddle $g_*$ used in \eqref{eq:fixednormstat}.  A straightforward computation \cite{Dong:2016fnf} then finds $I[g_n(A_*)] = nI[g_*] + (n-1)\frac{A_*}{4G}$, and thus $S_n = \frac{A_*}{4G}$.  In particular, the Renyi entropies $S_n(A_*)$ are independent of $n$ .  However, as usual, if $\partial R\neq \emptyset$ the Renyis diverge and require either a cutoff (say, defined using a certain boundary conformal frame) or renormalization to give finite results.

In general, one expects the RT area $A_{\gamma_R}$ to define superselection sectors of the quantum error correcting code associated with CFT reconstruction of the bulk entanglement wedges of $R$ and $\bar{R}$ \cite{Harlow:2016vwg}. When this is the case, the density matrix on $R$ of a CFT state $|\psi\rangle$ is block-diagonal $A_{\gamma_R}$, so that
\begin{equation}
\label{eq:block}
\rho=\oplus_{A_*} P(A_*) \rho_{A_*} ,
\end{equation}
with $\rho_{A_*}$ given by \eqref{rhoA0} in terms of the corresponding fixed-area state.  The representation \eqref{eq:block} motivates the idea that fixed-area states may be useful in studying the entropy of $|\psi\rangle$.  However, the arguments for \eqref{eq:block} (see \cite{Harlow:2016vwg}) are based (in part via \cite{Jafferis:2015del,Dong:2016eik}) on the Faulkner-Lewkowycz-Maldacena result \cite{Faulkner:2013ana} that the leading correction to $A/4G$ is of order $G^0$ and is given by bulk entanglement.  As described above, we expect this to fail near an RT phase transition\footnote{Such a failure is natural as \cite{Faulkner:2013ana} builds on the semi-classical Lewkowycz-Maldacena argument \cite{Lewkowycz:2013nqa}, which assumes a single RT surface to dominate.  This assumption clearly breaks down at an RT phase transition, and it is known that a proper treatment of cases with multiple extremal surfaces will be subtle; see e.g. comments in \cite{Fischetti:2014zja} based on a talk by Matt Headrick, which was in turn based on private remarks by Rob Myers.}.  So while \eqref{eq:block} may provide some motivation, we should take care not to rely on it to hold exactly in the regime of interest.

We conclude this section with a remark about notation.  Most of the explicit computations in sections \ref{Sec:SingleInterval} and \ref{Sec:Examples} will be for 3-dimensional bulk spacetimes.  In such cases codimension-2 extremal surfaces are geodesics and the associated `areas' are in fact lengths.  We will thus introduce $L_*=A_*$ and write all equations in those sections in terms of $L_*$, referring to it as the fixed length of the RT surface.  Once the reader is aware of this convention, it should create no confusion.  We will also generally drop the subscript $*$ below.

\section{Corrections to holographic entanglement entropy near phase transitions}\label{Sec:Correction}

We now turn to our main task of studying entropies of holographic states near RT phase transitions.  In particular, let us suppose our holographic state $|\psi \rangle$ is associated with a semi-classical geometry $g$ having two candidate RT surfaces $\gamma_1,\gamma_2$ associated with some partial Cauchy surface $R$ of the boundary spacetime.  Thus $\gamma_1,\gamma_2$ are both extremal surfaces anchored to the boundary $\partial R$ of $R$, and both are homologous to $R$ in the sense of \cite{Headrick:2007km}. Since our state is assumed to be pure, the surfaces $\gamma_1,\gamma_2$ are homologous to $\bar R$ as well.

We will proceed by considering a holographic state $|\psi\rangle$ and fixing the areas of both $\gamma_1$ and $\gamma_2$.
The probabilities $P(A_1,A_2)$ to obtain areas $A_1$ and $A_2$ can then be computed in direct analogy to the method described in section \ref{Sec:Review} for fixing the area of an RT surface.  In particular, we have
\begin{equation}
P(A_1,A_2) = \frac{{}_{A_1,A_2}\langle \psi | \psi \rangle_{A_1,A_2}}{\langle \psi | \psi \rangle},
\end{equation}
with ${\langle \psi | \psi \rangle} = e^{I + O(G^0)}$ and  ${}_{A_1,A_2}\langle \psi | \psi \rangle_{A_1,A_2}= e^{I(A_1,A_2) + O(G^0)}$ in terms of the Euclidean actions $I$, $I(A_1,A_2)$ of the leading saddles defined respectively by the path integral for
$\langle \psi | \psi \rangle$ and by the corresponding path integral with the areas of $\gamma_1, \gamma_2$ fixed to take the values $A_1,A_2$.  Recall that in the latter case the action generally includes a delta-function curvature contribution from both surfaces $\gamma_1$ and $\gamma_2$.  As before, the most likely values $\bar A_1, \bar A_2$ for our areas are just the values in the smooth saddle $g_0$ that dominates the path integral  for the norm $\langle \psi | \psi \rangle$ (and with no a priori fixing of areas).

Below, we first describe some of the topological details of our setup that will prove useful in the main argument.  We then discuss and motivate our diagonal approximation before computing the resulting $O(G^{-1/2})$ correction in section \ref{Sec:form}.

\subsection{Topological remarks}

For convenience  we will assume that while $\gamma_1$ and $\gamma_2$ are homologous, the two surfaces lie in distinct {\it homotopy} classes\footnote{Recall that homotopy is a more fine-grained equivalence relation than homology.}, and that each is the minimal-area such extremal surface within its homotopy class.  Having a topological distinction between the surfaces provides a natural definition of what we mean by the corresponding extremal surfaces $\gamma_1,\gamma_2$ in the conically-singular spacetimes associated with fixing the area of these extremal surfaces\footnote{This is merely a matter of convenience.  One could alternatively simply consider all saddle-points of the fixed-area action described in \cite{Dong:2019piw}, which describe spacetimes with what one may call extremal codimension-2 conical defects anchored to $\partial R$. It is not strictly necessary to label such conical defects as being associated with one of the extremal surfaces $\gamma_1,\gamma_2$ in the original smooth spacetime.}.  Furthermore, we will assume that -- at least for small defect angles and near the phase transition -- in all other homotopy classes the minimal surface $\gamma$ has area strictly greater than either $\gamma_1$ or $\gamma_2$.  This allows us to neglect such additional candidate RT surfaces in the semi-classical approximation.

Even in Lorentz signature, two extremal surfaces anchored on the same boundary set $\partial R$ are spacelike separated in the bulk and lie on a common Cauchy surface $\Sigma$ \cite{Wall:2012uf}.  We note that this is the case even when $\partial R = \emptyset$. As a result, the associated RT area operators $\hat A_1$, $\hat A_2$ for $\gamma_1,\gamma_2$ commute at all orders in the semi-classical expansion and -- at least at this level --  can be simultaneously diagonalized.  In particular, the possible obstruction described in \cite{Bao:2018pvs} does not arise.  We may thus consider the doubly-fixed-area states $|\psi\rangle_{A_1,A_2}$ in which the area of $\gamma_1$ is $A_1$ and the area of $\gamma_2$ is $A_2$.  Here we introduce an appropriate UV cutoff in the boundary to render $A_1,A_2$ finite.    Since both are anchored on the same set $\partial R$, we use the same cutoff to define both $A_1$ and $A_2$.

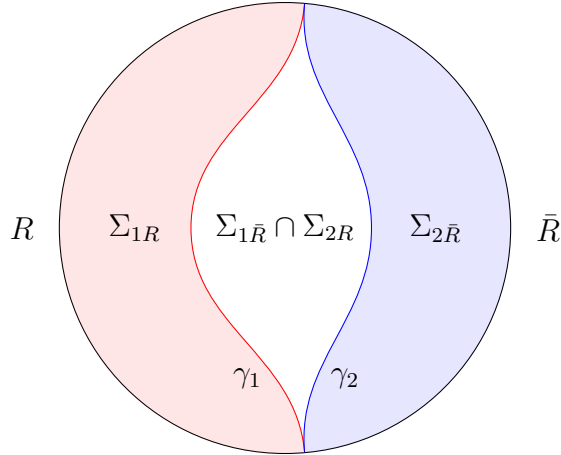
\begin{figure}
\centering
\begin{tikzpicture}
\draw (0,0) circle [radius=3];
\draw [name path=B,red] (0.26,2.98) to [out=265,in=90] (-1.25,0) to [out=-90,in=95] (0.26,-2.98) ;
\path [name path=A] (0,3) arc (90:270:3);
\draw [name path=C,blue] (0.26,2.98) to [out=265,in=90] (1.15,0) to [out=-90,in=95] (0.26,-2.98);
\path [name path=D] (0,3) arc (90:-90:3);
\tikzfillbetween[of=A and B]{red, opacity=0.1};
\tikzfillbetween[of=C and D]{blue, opacity=0.1};
\node at (-3.5,0) {$R$};
\node at (3.5,0) {$\bar{R}$};
\node at (-0.5,-2) {$\gamma_1$};
\node at (0.8,-2) {$\gamma_2$};
\node at (-2,0) {$\Sigma_{1R}$};
\node at (2,0) {$\Sigma_{2\bar{R}}$};
\node at (0,0) {$\Sigma_{1\bar R} \cap \Sigma_{2R}$};
\end{tikzpicture}
\caption{An illustration of two competing RT surfaces $\gamma_1$ and $\gamma_2$ near a phase transition. In our convention, we always let $\Sigma_{1R}\subset \Sigma_{2R}$, as a result $\Sigma_{1R}$ and $\Sigma_{2\bar{R}}$ are non-overlapping. }
\label{fig:2gammas}
\end{figure}

The homology constraint means that each surface $\gamma_{i}$  ($i \in \{1,2\}$) must partition $\Sigma$ into two (non-overlapping) parts $\Sigma_{iR}, \Sigma_{i\bar R}$ where $\partial \Sigma_{iR} = \gamma_i \cup R$ and similarly for $\partial \Sigma_{i\bar R}$; see figure \ref{fig:2gammas}.  We will further assume that $\Sigma_{1R}$ is contained in $\Sigma_{2R}$.  At least in the time-symmetric case, this assumption can be made without loss of generality.  To see this, note that we must have either $\Sigma_{1R} \subset \Sigma_{2R}$, $\Sigma_{2R} \subset \Sigma_{1R}$, or that $\gamma_2$ enters both $\Sigma_{1R}$ and $\Sigma_{1\bar R}$. The first case fulfills our assumption, and in the second case the assumption can be fulfilled by simply exchanging the labels $1 \leftrightarrow 2$.

In the third case, the intersection $\gamma_{int} = \gamma_1 \cup \gamma_2$ partitions $\gamma_2$ into two parts $\gamma_{2R}\subset \Sigma_{1R}$ and $\gamma_{2\bar R} \subset \Sigma_{1\bar R}$. Similarly, we must also find that $\gamma_1$ enters both $\Sigma_{2R}$ and $\Sigma_{2\bar R}$, so $\gamma_{int}$ also partitions  $\gamma_1$ into two parts $\gamma_{1R} \subset \Sigma_{2R}$ and $\gamma_{1\bar R} \subset \Sigma_{2\bar R}$.  Note that $\gamma_{1R}$ and $\gamma_{2\bar R}$ must be homologous but cannot be homotopic. Similarly, $\gamma_{2R}$ and $\gamma_{1\bar R}$ must be homologous but cannot be homotopic.

For this case, let us choose the labels $1$ and $2$ so that $\gamma_{2R}$ has smaller area than $\gamma_{1\bar R}$ and define a new surface $\gamma_3 = \gamma_{1R} \cup \gamma_{2 R}$.  Note that $\gamma_3$ also satisfies the homology constraint, but that it cannot be homotopic to either $\gamma_1$ or $\gamma_2$.  While $\gamma_3$ is not extremal, it has area $A_3$ satisfying $A_3 < A_1$.  So the minimal surface within its homotopy class also has area less than $A_1$.  But this contradicts the earlier assumption that the least-area extremal surface in any other homotopy class must have area strictly greater than either $A_1$ or $A_2$.  Thus our 3rd case cannot exist in the time-symmetric case, and we can take $\Sigma_{1R} \subset \Sigma_{2R}$ without loss of generality\footnote{It would be interesting to understand if this result continues to hold without time symmetry.  If it does, the rest of the argument generalizes in a straightforward way to the HRT case using the Schwinger-Keldysh techniques of \cite{Dong:2016hjy}.}.

\subsection{The diagonal approximation}
\label{Sec:Diag}

Because the states $|\psi\rangle_{A_1,A_2}$ are at least approximate eigenstates of $\hat A_1,\hat A_2$, any two such states are semi-classically orthogonal unless they have fixed the same values for the areas of both $\gamma_1$ and $\gamma_2$.
The fixed-area states thus naturally define a decomposition of $|\psi\rangle$ according to
\begin{equation}
\label{eq:psi}
|\psi \rangle  = \sum_{A_1,A_2} \sqrt{\frac{P(A_1,A_2)}{\langle \psi|\psi \rangle}} |\psi \rangle_{A_1,A_2}.
\end{equation}
As in section \ref{Sec:Review}, we take the states ${|\psi \rangle}_{A_1,A_2}$ to be associated with finite intervals of $A_1,A_2$ that are small with respect to the semiclassical width of $P(A_1,A_2)$ but large compared with the spacing between adjacent area eigenstates. We thus take the intervals to be polynomially small in $G$, but not exponentially small.

It now remains to compute the density matrix $\rho_R$ on the region $R$ by tracing $|\psi\rangle \langle \psi |$ over the complementary region $\bar R$:
\begin{equation}
\rho_R = \sum_{A_1,A_2,A_1{}',A_2{}'} \sqrt{P(A_1,A_2)}\sqrt{P(A_1{}',A_2{}')} \rm{Tr}_{\bar R} \left( \frac { {|\psi \rangle}_{A_1,A_2} {}_{A_1{}',A_2{}'}{\langle \psi|}}{\langle \psi | \psi \rangle}\right).
\end{equation}
In doing so, one must consider contributions from both diagonal terms (with $A_1 = A_1{}'$ and $A_2 = A_2{}'$) as well as contributions from off-diagonal terms (where either $A_1 \neq A_1{}'$ or $A_2 \neq A_2{}'$).

The diagonal terms give the average over the distribution $P(A_1,A_2)$ of the (normalized) density matrices $\rho_R(A_1,A_2)$ defined by the normalized fixed-area states.  Let us therefore write
\begin{equation}
\label{eq:rhoRdecomp}
\rho_R = \sum_{A_1,A_2} P(A_1,A_2) \rho_R(A_1,A_2) + OD_R,
\end{equation}
where $OD_R$ is the result of summing all off-diagonal contributions.

Since $A_1$ defines a Hermitian operator that can be reconstructed on $R$, we must have
$ \rho_R(A_1,A_2) \rho_R(A_1{}',A_2{}') =0$ for $A_1\neq A_1{}'$.  Note that the same need not always hold for $A_2$ since it can be reconstructed on $R$ only for $A_2 < A_1$.  However, if we instead considered the density matrices on $\bar R$  that result from tracing over $R$, this would interchange the roles of $A_1$ and $A_2$, suggesting that the full problem exhibits a greater symmetry. We will therefore treat the $\rho_R(A_1,A_2)$ below as if they live in orthogonal subspaces\footnote{We emphasize that this is an additional assumption and thank Geoffrey Penington for conversations related to this point.  \label{foot:add}}.

Let us first discuss the contributions of the diagonal terms.  In particular, we introduce the von Neumann entropies
\begin{equation}
\label{eq:rhoD}
S_D = - {\rm Tr} \left( \rho_D \ln \rho_D \right) \ \ \ \ {\rm for} \ \ \ \ \rho_D := \oplus_{A_1,A_2} P(A_1,A_2) \rho_R(A_1,A_2),
\end{equation}
\begin{equation}
S(A_1,A_2) = - {\rm Tr} \left( \rho_R(A_1,A_2) \ln \rho_R(A_1,A_2) \right).
\end{equation}
Treating the diagonal terms as living in orthogonal subspaces, a standard computation shows these quantities to be related by
\begin{equation}
\label{eq:Diagansw}
S_D = \sum_{A_1,A_2} \left( P(A_1,A_2) S(A_1,A_2) - P(A_1,A_2) \ln P(A_1,A_2)) \right),
\end{equation}
where the 2nd term is often called the entropy of mixing.  The entropy of mixing is bounded by the logarithm of the number of values that the pair $(A_1,A_2)$ can take.  Since each value $(A_1,A_2)$ labels an interval that is only polynomially small in $G$, this bound is of the form $C \ln G + s(\psi)$ where $C$ is an order-one constant and dependence on the state $\psi$ appears only through the order-one function $s(\psi)$. We will thus neglect the entropy of mixing below since it is parametrically smaller than the $O(G^{-1/2})$ term we wish to study.

Now, before returning to the off-diagonal terms $OD_R$, we also wish to compute $S(A_1,A_2)$.  As reviewed in section \ref{Sec:Review}, the fact that fixed-area states allow arbitrary conical singularities at the associated extremal surfaces means that the semiclassical Renyi entropies of such states are straightforward to compute. In particular, every $n$-sheeted branched cover of the original Euclidean geometry defines a saddle for the $n$th Renyi problem.  Furthermore, comparison with tensor networks suggests that all Renyi saddles are of this form.

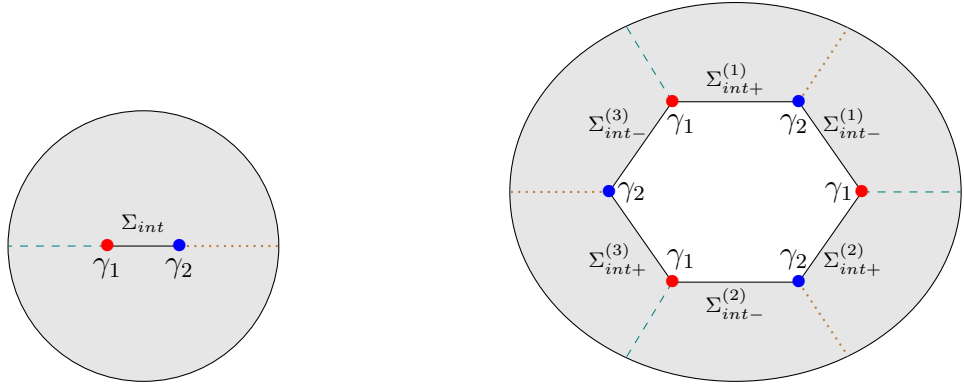
\begin{figure}
\begin{minipage}[t]{0.5\textwidth}
\centering
\begin{tikzpicture}[scale=0.6]
\draw [fill=gray!20] (0,0) circle [radius=3];
\draw (-0.8,0) to (0.8,0);
\draw [dashed,teal] (-0.8,0) to (-3,0);
\draw [dotted,brown,thick] (0.8,0) to (3,0);
\node [red] at (-0.8,0) {$\bullet$};
\node [blue] at (0.8,0) {$\bullet$};
\node at (-0.8,-0.5) {$\gamma_1$};
\node at (0.8,-0.5) {$\gamma_2$};
\node (c) at (0,0.5) {\fontsize{8}{8}\selectfont $\Sigma_{int}$};
\end{tikzpicture}
\end{minipage}
\begin{minipage}[t]{0.5\textwidth}
\centering
\begin{tikzpicture}[scale=0.6]
\draw [fill=gray!20] (0,0) ellipse (5 and 4.2);
\draw [fill=white] (-1.4,-2) to (-2.8,0) to (-1.4,2) to (1.4,2) to (2.8,0) to (1.4,-2) to (-1.4,-2);
\draw [dashed,teal] (-1.4,-2) to (-2.5,-3.8);
\draw [dashed,teal] (-1.4,2) to (-2.5,3.8);
\draw [dashed,teal] (2.8,0) to (5,0);
\draw [dotted,brown,thick] (1.4,2) to (2.5,3.65);
\draw [dotted,brown,thick] (1.4,-2) to (2.5,-3.65);
\draw [dotted,brown,thick] (-2.8,0) to (-5,0);
\node [red] at (-1.4,-2) {$\bullet$};
\node [red] at (-1.4,2) {$\bullet$};
\node [red] at (2.8,0) {$\bullet$};
\node [blue] at (-2.8,0) {$\bullet$};
\node [blue] at (1.4,2) {$\bullet$};
\node [blue] at (1.4,-2) {$\bullet$};
\node at (-1.2,1.5) {$\gamma_1$};
\node at (-1.2,-1.5) {$\gamma_1$};
\node at (2.3,0) {$\gamma_1$};
\node at (1.3,1.5) {$\gamma_2$};
\node at (1.3,-1.5) {$\gamma_2$};
\node at (-2.3,0) {$\gamma_2$};
\node at (-2.6,1.5) {\fontsize{8}{8}\selectfont $\Sigma_{int-}^{(3)}$};
\node at (-2.6,-1.5) {\fontsize{8}{8}\selectfont $\Sigma_{int+}^{(3)}$};
\node at (0,2.5) {\fontsize{8}{8}\selectfont $\Sigma_{int+}^{(1)}$};
\node at (2.6,1.5) {\fontsize{8}{8}\selectfont $\Sigma_{int-}^{(1)}$};
\node at (2.6,-1.5) {\fontsize{8}{8}\selectfont $\Sigma_{int+}^{(2)}$};
\node at (0,-2.5) {\fontsize{8}{8}\selectfont $\Sigma_{int-}^{(2)}$};

\end{tikzpicture}
\end{minipage}
\caption{ \textbf{Left:}
 A two-dimensional projection of an $n=1$ solution with two extremal surfaces $\gamma_1,\gamma_2$ (having areas $A_1$ and $A_2$) and a surface $\Sigma_{int}$ stretching between them.
  \textbf{Right:} An $n$-fold cover of the figure at left for the case $n=3$ after cutting open a slit along $\Sigma_{int}$.  The $2n$ copies of $\Sigma_{int}$ are labeled $\Sigma^{(i)}_{int\pm}$, where $i=1,\dots,n$.
  Saddles for the R\'enyi entropy are formed by identifying $\Sigma^{(i)}_{int+}$ with $\Sigma^{(\pi(i))}_{int-}$ for some permutation $\pi$.
   After making such identifications, the number $N_2 = n-n_2$  of copies of $\gamma_2$ that remain is the number $C(\pi)$ of cycles generated by $\pi$, while the corresponding $N_1 = n-n_1$ is $C(\tau \circ \pi)$ where $\tau$ is the counterclockwise cyclic permutation.    The asymmetry is due to the numbering of replicas, which breaks the natural symmetry between the dashed lines (separating replicas) and dotted lines (separating the two halves of each replica).}\label{fig:RBsaddles}
\end{figure}

In our present case, the branching can occur at either surface $\gamma_1$ or $\gamma_2$, or on any of their Renyi copies.
Note that the surfaces $\gamma_1,\gamma_2$ partition the time-symmetric surface $\Sigma$ into 3 parts according to $\Sigma = \Sigma_{1R} \cup \Sigma_{int} \cup \Sigma_{2\bar R}$ where $\Sigma_{int} = \Sigma_{1\bar R} \cap \Sigma_{2R}$ lies between $\gamma_1$ and $\gamma_2$.
The possible saddles can then be constructed by the following procedure.  First, cut a slit along $\Sigma_{int}$  in the original spacetime $g_0$ to define a spacetime with an internal boundary $\Sigma_{int+} \cup \Sigma_{int-}$, where $\Sigma_{int\pm}$ are the two sides of the newly-opened slit along $\Sigma_{int}$.  Next consider the $n$-fold cover of the result that winds $n$ times around this slit; see figure \ref{fig:RBsaddles}.  Finally, sew the up the slit by making identifications between the $n$ copies of $\Sigma_{int+}$ and the $n$ copies of $\Sigma_{int-}$.  Since there are $n! = \Gamma(n+1)$ ways to pair up the copies of $\Sigma_{int+}$ and $\Sigma_{int-}$, this results in $\Gamma(n+1)$ saddles.

However, as shown in \cite{Dong:2018seb} the fixed-area action of a branched cover depends only on the action of the spacetime $g_{0,0}$ that dominates the fixed-area path integral for ${}_{A_1,A_2} \langle \psi | \psi\rangle _{A_1,A_2}$ and on the conical defects and areas of the branching surfaces.  As a result, for a given branched-cover the Euclidean action depends only on the numbers $n_1,n_2$ of times that it branches over each of $\gamma_1,\gamma_2$, irrespective of the order in which those branchings occur.  In more detail, we take $2\pi n_1$ to be the sum of the conical excesses around all copies of $\gamma_1$, and similarly for $n_2$.

The action for these saddles follows from the analysis of \cite{Dong:2018seb}, which yields
\begin{equation}
\label{eq:branchaction}
I[g_{n_1,n_2}] = nI[g_0] + \frac{n_1A_1 + n_2 A_2}{4G}.
\end{equation}
Here\footnote{We thank Geoffrey Penington for pointing out an error in a previous draft and for conversations related to the comments below.} $n_1 + n_2 \ge n-1$,  consistent with the fact that no branching occurs for $n=1$. To minimize the action, we will be interested in saddles that saturate this inequality (so that $n_1 +n_2 = n-1$).  To understand these saddles, we may describe the above sewing by a permutation $\pi$ of the copies of $\Sigma_{int-}$ relative to the copies of $\Sigma_{int+}$. As shown in figure \ref{fig:RBsaddles}, any numbering of the copies of $\Sigma_{int\pm}$ breaks a natural symmetry between $\gamma_1$ and $\gamma_2$.  As a result, with the conventions of figure \ref{fig:RBsaddles}, the number $N_2$ of copies of $\gamma_2$ that remain after these identifications is given by the number $C(\pi)$ of closed cycles  associated with the permutation $\pi$ (e.g., the permutation $(12)(3)$ on 3 objects has $C=2$), while the corresponding $N_1$ is given by the number $C(\tau \circ \pi)$ where $\tau$ is the cyclic permutation that maps copy $i$ to copy $i-1$).  Since the winding numbers $n_1,n_2$ defined above are $n_1 = n-N_1$, $n_2= n-N_2$, we have $N_1 + N_2 = 2n - (n_1 + n_2) \le n+1$ and we wish to saturate this bound.
As described in appendix E of \cite{Penington:2019kki}, the number of permutations on $n$ objects that do so (and  thus which have $N_1 + N_2 = C(\pi) + C(\pi \circ \tau)=n+1$) is given by the $n$th Catalan number $C_n = \frac{1}{n+1}{2n \choose n} = \frac{\Gamma(2n+1)}{(n+1)[\Gamma(n+1)]^2}$.

When $A_1$ and $A_2$ differ significantly, the $n$th Renyi is clearly dominated by a saddle with action $I =  nI[g_0] + \frac{n-1}{4G} \rm{min}(A_1,A_2)$ and we find $S(A_1,A_2) = \frac{\rm{min}(A_1,A_2)}{4G}$ in direct analogy with the case studied in \cite{Dong:2018seb} where only one area is fixed.  On the other hand, when $A_1 = A_2$ all of the $C_n$ saddles with $n_1+n_2=n-1$ have the same action\footnote{Because the number $\Gamma(n+1)-C_n$ of other saddles vanishes at $n=1$, the other saddles can contribute at most an $O(1)$ correction to the von Neumann entropy.  That is enough for us to drop such contributions.  But the interested reader can find more discussion in \cite{Penington:2019kki}, and it appears that the contribution of such saddles to the von Neumann entropy is in fact non-perturbatively small, being proportional to $e^{-\frac{A_{1,2}}{4G}}$ and thus vanishing exactly when $\gamma_{1,2}$ reach the boundary and  $A_{1,2}$ diverge.} $I[g_{n-1,0}]$ and we find
\begin{equation}
S_n = \frac{1}{1-n}\ln \frac{Z_n}{Z_1^n} \approx \frac{ \ln C_n + I[g_{n-1,0}] -nI[g_0] }{1-n} = \frac{A_1}{4G} - \frac{1}{n-1}\ln \left(\frac{\Gamma(2n+1)}{(n+1)[\Gamma(n+1)]^2}\right).
\end{equation}
in terms of the $n$th Renyi partitions functions $Z_n$. Note that the final term is of order $G^0$ and has a finite limit $ -2\frac{\Gamma'(3)}{\Gamma(3)} + \frac{1}{2} + 2\frac{\Gamma'(2)}{\Gamma(2)}=-\frac{1}{2}$ as $n \rightarrow 1$.  Since it is clear that the largest correction will occur for $A_1 = A_2$, we may thus write
$S(A_1,A_2) = \frac{\rm{min}(A_1,A_2)}{4G} + O(G^0)$ for all $A_1,A_2$.

\begin{figure}
\centering
\begin{tikzpicture}[scale=0.6]
\draw [fill=gray!20] (2.95,0.52) arc (10:350:3);
\path [fill=white] (2.95,0.52) to (2.95,-0.52) to (0.5,0);
\draw [thick,green] (0.5,0) to (2.95,0.52);
\draw [thick,orange] (0.5,0) to (2.95,-0.52);
\node [red] at (-0.5,0) {$\bullet$};
\node [blue] at (0.5,0) {$\bullet$};
\node at (-0.5,-0.5) {$A_1$};
\node at (0.5,-0.5) {$A_2$};
\end{tikzpicture}
\begin{tikzpicture}[scale=0.6]
\draw [fill=gray!20] (2.95,0.52) arc (10:350:3);
\path [fill=white] (2.95,0.52) to (2.95,-0.52) to (0.5,0);
\draw [thick,green] (0.5,0) to (2.95,-0.52);
\draw [thick,orange] (0.5,0) to (2.95,0.52);
\node [red] at (-0.5,0) {$\bullet$};
\node [blue] at (0.5,0) {$\bullet$};
\node at (-0.5,-0.5) {$A'_1$};
\node at (0.5,-0.5) {$A_2$};
\end{tikzpicture}
\caption{ Using the same projection as for the $n=1$ figure at left in figure \ref{fig:RBsaddles}, we show two pieces of a corresponding saddle that for $A_1\neq A'_1$  describes an off-diagonal contribution to the second R\'enyi entropy $S_2$. The full saddle is constructed by sewing the two pieces together along edges of the same color; i.e., we may first identify the two green edges and then identify the two orange edges. Note that this particular saddle contains only one copy of the surface $\gamma_2$ and so cannot be `off-diagonal' in $A_2$.}
\label{fig:offdiag}
\end{figure}
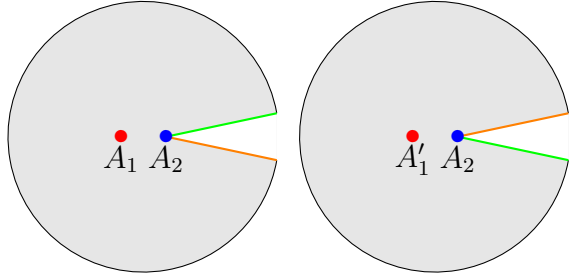

We now return to the off-diagonal term $OD_R$ in \eqref{eq:rhoRdecomp} and its contributions to the Renyi entropies $S_n(\rho_R)$. While we leave full consideration of such terms to future work, we will give a plausibility argument suggesting that these contributions can be ignored for our current purposes.
To begin this plausibility argument note that,
in the semiclassical approximation, each such contribution can be written as $e^{-I}$ where $I$ is the action of a branched cover of $g_0$ similar to those described above, but where the areas of the various Renyi copies of $\gamma_1$ can differ from each other\footnote{We thank Xi Dong, Geoffrey Penington, Xiaoliang Qi, and  Douglas Stanford for discussions of this point.}, and similarly for the Renyi copies of $\gamma_2$. See figure \ref{fig:offdiag}. In particular, at least at the leading semiclassical order discussed here, such contributions are associated with the possibility of breaking replica symmetry.  Since a strict breaking of replica symmetry is impossible at $n=1$, it is plausible that their contribution will be subleading in the limit where the replica number $n$ is taken to $1$.
In particular, since for any normalized $\rho_R$ the diagonal terms yield $S_{n,{\rm diag}} = O(n-1)$, it is plausible that off diagonal contributions will be of order $O\left((n-1)^2\right)$ or of order $G^0$ (from corrections to the leading semiclassical terms).  For now, we simply assume that this is the case and follow up by checking consistency with results from \cite{Murthy:2019qvb} and \cite{Penington:2019kki} in section \ref{Sec:Examples}.

With the above assumption, the von Neumann entropy $S(\rho_R)$ is given by just the diagonal contributions $S_D(\rho_R)$ up to corrections of order $G^0$ and, introducing a normalization factor $N$, we write
\begin{align}
\label{eq:diagdom}
S(\rho_R) := S_D + O(G^0) &=  \sum_{A_1,A_2}  \frac{\rm{min}(A_1,A_2)}{4G} P(A_1,A_2) + O(G^0) \cr
&=  \sum_{A_1,A_2}  \frac{\rm{min}(A_1,A_2)}{4G} N e^{-(I(A_1,A_2) - I(\bar A_1,\bar A_2))} + O(G^0) \cr
&= \int dA_1dA_2  \frac{\rm{min}(A_1,A_2)}{4G} N e^{-(I(A_1,A_2) - I(\bar A_1,\bar A_2))} + O(G^0), \ \ \ \ \ \
\end{align}
where in the last step we may approximate the sum by an integral with error smaller than $O(G^0)$ since the spacings between values of $A_1,A_2$ included in the sum were taken to be small compared with the natural scale of variation of $I(A_1,A_2)$.

\subsection{The general form of corrections}
\label{Sec:form}

The above section motivated the diagonal approximation \eqref{eq:diagdom} $S(\rho_R) := S_D + O(G^0)$ (with $S_D$ given by \eqref{eq:Diagansw}) and derived the resulting simple form \eqref{eq:diagdom} for $S(\rho_R)$ in terms of the fixed-area actions.  We will now show how this form gives an $O(G^{-1/2})$ correction to the RT entropy. This merely requires evaluating the final integral in \eqref{eq:diagdom} in the semiclassical limit $G \rightarrow 0$.

Since the action $I$ is proportional to $1/G$,  taking $G\rightarrow 0$ concentrates the integral near the areas $\bar A_1, \bar A_2$ that minimize the action.  As usual, we can approximate $I$ near that minimum as quadratic:
\begin{equation}
\label{eq:quad}
I(A_1,A_2) = I(\bar A_1, \bar A_2) + \frac{1}{2}\sum_{i,j=1}^{2} \left(\frac{\partial^2 I}{\partial A_i\partial A_j}\Big|_{\bar A_i} \right) (A_i-\bar A_i) (A_j-\bar A_j) + O\left((A-\bar A)^3\right),
\end{equation}
where the cubic and higher terms in \eqref{eq:quad} will contribute to \eqref{eq:diagdom} only at order $G^0$.  We may neglect all such terms since our goal is to study corrections at order $G^{-1/2}$.  It will prove useful in analyzing the examples of section \ref{Sec:Examples} that \eqref{eq:quad} involves only configurations infinitesimally close to the smooth saddle $g_{0,0}$ that dominates the path integral for $\langle \psi |\psi\rangle$, and in particular which has vanishing cosmic brane tension (and thus vanishing conical defect) on both $\gamma_1$ and $\gamma_2$.

For later reference, we introduce the covariance matrix $C$ with components
\begin{equation}
\label{eq:cov}
C_{ij} =
\begin{bmatrix}\sigma_1^2&r\sigma_1\sigma_2\\r\sigma_1\sigma_2&\sigma_2^2\\\end{bmatrix}=
G \begin{bmatrix}\tilde \sigma_1^2&r\tilde \sigma_1 \tilde \sigma_2\\r\tilde \sigma_1\tilde \sigma_2&\tilde \sigma_2^2\\\end{bmatrix}
\end{equation}
defined by $(C^{-1})_{ij} = \frac{\partial^2 I}{\partial A_i\partial A_j}\Big|_{\bar A_i}.$  Note that since $\frac{\partial^2 I}{\partial A_i\partial A_j}$ is of order $1/G$, the covariance matrix is of order $G$.   The final form in \eqref{eq:cov} displays this $G$-dependence explicitly, and the parameters $\tilde \sigma_1, \tilde \sigma_2, r$ are all of order $G^0$.

The integral defined by using \eqref{eq:quad} in \eqref{eq:diagdom} is naturally studied in terms of the variables $A_\pm = \frac{A_1 \pm A_2}{2}$ for which we have ${\rm min}(A_1,A_2) = A_+ - |A_-|$ and the most likely values are $\bar A_\pm := \frac{\bar A_1 \pm \bar A_2}{2}$.  The integral over $A_+$ is straightforward and gives
\begin{equation}
S_D = \int^{\infty}_{-\infty} dA_- \frac{(A_- -\bar A_-)(\tilde \sigma_1^2 - \tilde \sigma^2_2) +  4 (\bar A_+ - |A_-|)  \tilde \sigma_-^2}
{16G^{3/2} \sqrt{2\pi} \tilde \sigma_-^3}\exp {\left( -\frac{(A_- - \bar A_-)^2}{2G \tilde \sigma_-^2 }\right)} + O(G^0),
\end{equation}
where $4\tilde \sigma_-^2 = \tilde \sigma_1^2-2r\tilde \sigma_1\tilde \sigma_2+\tilde  \sigma_2^2$.  We will also use $\sigma_-^2 = G \tilde \sigma_-^2$ below.

The term linear in $(A_- -\bar A_-)$ integrates to zero by symmetry.  The remaining integral can be written in terms of the error function $\erf x : = \frac{2}{\sqrt{\pi}}\int_0^x dt e^{-t^2}$ as
\begin{equation}
\label{eq:entcor}
\begin{split}
S(\rho) &= \frac{\bar A_{+}}{4G} - \frac{\bar A_{-}}{4G} \erf{\left( \frac{\bar A_{-}}{\sqrt{2G} \tilde \sigma_- }\right)}
- \frac{\tilde \sigma_-}{2\sqrt{2\pi G}}\exp{\left(-\frac{\bar{A}_-^2 }{2G \tilde \sigma_-^2}\right)}
 + O(G^0)
 \\
& =
\frac{\min\left( \bar A_1, \bar A_2 \right) }{4G}  -  \frac{\tilde \sigma_-}{2\sqrt{2\pi G}} \Phi\left(\frac{\bar{A}_1 - \bar A_2 }{2 \tilde \sigma_- \sqrt{2G} }\right)  +
 O(G^0),
\end{split}
\end{equation}
where we have introduced
\begin{equation}
\label{eq:Phi}
\Phi(x):= e^{-x^2} + \sqrt{\pi} |x| (\erf |x|-1 )
\end{equation}
following the notation of \cite{Murthy:2019qvb}. Note that $\Phi(x)$ is bounded by a constant of order $G^0$,
The final expression in \eqref{eq:entcor} thus makes manifest that we find a correction of order $G^{-1/2}$ at the transition where $\bar A_1 = \bar A_2$, but that the correction is  exponentially small at large $|A_1-A_2|/\tilde \sigma_-\sqrt{G} = |A_1-A_2|/\sigma_-$.  On the other hand, the first line in \eqref{eq:entcor} shows that the entropy at this order is a smooth function of $\bar A_1 - \bar A_2$; the supposed RT `phase transition' is in fact already a crossover at this level of analysis.

\section{Fixed length states for a single interval in the AdS$_3$ vacuum}
\label{Sec:SingleInterval}
We now wish to perform explicit computations illustrating the above $O(G^{-1/2})$ correction and exploring the size of fluctuations in RT-areas in various examples.  However, before doing so it is useful to analyze fixed-area states associated with a simple case in which phase transition do {\it not} arise.  We do so here, studying the particularly simple case where we choose our boundary region $R$ to be a single interval on the $t=0$ slice of the boundary of empty global AdS$_3$. Since two of our examples in section \ref{Sec:Examples} below will also involve intervals on the boundary of either AdS$_3$ or a BTZ quotient, we will be able to use results obtained below to simplify the analysis of those phase transitions.

As usual, in order to find the probability that the RT surface for our single-interval $R$ has some fixed length $L_*$, we will use the saddle-point approximation and study the action for the appropriate classical Euclidean solution.  As discussed above, this solution will have a conical defect (which in AdS$_3$ takes the form of a spacelike cosmic string).  For pure Einstein-Hilbert AdS$_3$ gravity, the fact that all solutions are locally equivalent to AdS$_3$ means that the solution with fixed length $L_*$ may be constructed from global AdS$_3$ by inserting a conical singularity along the associated RT surface and tuning the conical  angle so that the length becomes $L_*$ as defined by using an appropriate cutoff with respect to the desired conformal frame at infinity.

We thus begin by recalling that the Euclidean $AdS_3$ vacuum can be described as the Poincar\'e ball with metric
\begin{equation}
\label{Poincare}
ds^2=\dfrac{dr^2+r^2d\theta^2+ r^2\sin^2\theta d\phi^2}{(1-r^2/4)^2}.
\end{equation}
In \eqref{Poincare}, the coordinate ranges are $\theta \in [0,\pi]$, $\phi \in [0,2\pi)$, and $r \in [0,2)$. The AdS boundary lies at $r=2$ and we have set the bulk AdS length scale $\ell$ to $1$.

It is then straightforward to address the case where the boundary region $R$ is half of a great circle on the boundary $S^2$.  With an appropriate Wick rotation, we may thus think of this as half of the boundary circle at Lorentzian time $t=0$.  However, for our current purposes it will be convenient to take this interval to be the half-circle $\theta \in [0,\pi]$ at $\phi =0$; i.e., we take it to be the prime meridian instead of half of the equator.

By symmetry, the corresponding RT surface is then just the $\phi$-axis. Fixed length states for $R$ will then be associated with similar Euclidean solutions in which this axis is a conical singularity.  General such solutions are then described by inserting a positive factor $\alpha$ into \eqref{Poincare} to yield
\begin{equation}
\label{Poincare2}
ds^2=\dfrac{dr^2+r^2d\theta^2+\alpha^2 r^2\sin^2\theta d\phi^2}{(1-r^2/4)^2}.
\end{equation}

Note that we may also write \eqref{Poincare2} in terms of a rescaled angular coordinate $\tilde \phi = \alpha \phi$ with $\tilde \phi \in [0,2 \pi \alpha)$ to give a perhaps-more-familiar description of this conical metric.  The cases $\alpha <1$ describe conical deficits, while $\alpha >1$ is a conical excess.  Using the Einstein equations to interpret this conical defect as a (Euclidean) cosmic string, one finds that the string has a tension $\mu$ such that
\begin{equation}
\label{eq:alphamu}
\alpha=1-4\mu G.
\end{equation}
In particular, the string tension is negative for geometries with a conical excess.

We wish to fix the length of our defect cosmic string.   Of course, the actual length of the $\phi$-axis diverges but, as mentioned above we in fact wish to specify an appropriately regularized notion of its length.  We will do so by introducing a UV cutoff in the dual CFT, which then defines a radial cutoff in the bulk.  This requires specifying a conformal frame, and it is natural to take this to be the frame in which the boundary geometry is given by the round metric on the unit-radius $S^2$.

For $\alpha \neq 1$ this round conformal frame differs from the conformal frame naturally associated with the bulk metric \eqref{Poincare2}.  In particular, for $\alpha \neq 1$ multiplying \eqref{Poincare2} by $(1-r^2/4)^2$ and setting $r=2$ would give a boundary metric with conical singularities at both poles.  Of course, this conical metric is related to the round metric by a conformal (aka Weyl) rescaling of the metric.  Both the associated conformal factor $\Omega$ and the polar coordinate $\tilde \theta$ associated with the standard presentation of the round metric can be found by writing
\begin{equation}
d\theta^2+\alpha^2 \sin^2\theta d\phi^2=\Omega^2 (d\tilde{\theta}^2+\sin^2\tilde{\theta} d\phi^2).
\end{equation}
Solving for $\Omega^2$ and $\theta$ gives
\begin{equation}
\Omega^2=\left( \frac{\alpha \sin \theta(1+(\tan \frac{\theta}{2})^{2/\alpha} )}{2 (\tan \frac{\theta}{2})^{1/\alpha}}\right)^2
\end{equation}
and
\begin{equation}
\label{eq:thetas}
\theta=2 \tan^{-1}\left[ \left( \tan \frac{\tilde{\theta}}{2}\right)^\alpha \right].
\end{equation}

We take our UV cutoff to be given by a distance $\delta$ in the boundary as defined in the round unit-sphere conformal frame.  The associated bulk radial cutoff would then be at $z=\delta$ where $z$ is the Fefferman-Graham coordinate associated with the same round conformal frame.  However, for $\alpha \neq 1$ the conical singularity in \eqref{Poincare2} makes it non-trivial to write our metric in such coordinates.  So instead of explicitly computing the transformation between $r,\theta,\phi$ and the desired Fefferman-Graham coordinates, we will use the well-known fact that (to leading order in $\delta$) the desired cutoff $z=\delta$ can be identified as the greatest depth to which minimal surface anchored on a circle of size $\delta$ (as defined in the desired conformal frame) hangs down into the bulk.  In particular, since our conical singularity lies on the $\phi$-axis, it should be cutoff where it intersects the minimal surface anchored to a circle of round-frame radius $\delta$ about the pole $\tilde \theta =0$.  Note that the bulk conical singularity will prevent the minimal surface from being smooth, but that -- as is most easily seen for the case $\delta = \pi/4$ -- symmetry requires the surface to be invariant under an appropriate ${\mathbb Z}_2$ reflection.  This condition implies that the minimal surface must still intersect the axis orthogonally.

In the conical boundary-conformal frame, we see from \eqref{eq:thetas} that the surface is anchored at  $\theta=\delta_\alpha \equiv \tan^{-1} \left[ \left(\tan \frac{\delta}{2}\right)^\alpha \right]$.  A short computation shows that the
desired minimal surface satisfies
\begin{equation}
\dfrac{1}{r}+\dfrac{r}{4}=\dfrac{\cos\theta}{\cos \delta_\alpha}.
\end{equation}
The intersection with the $\theta =0$ axis occurs at $r_\alpha=2(\sec\delta_\alpha-\tan\delta_\alpha)$, so the cutoff RT surface (i.e., the cutoff cosmic string) has length
\begin{equation}
\label{eq:Clength}
\begin{aligned}
L &=2 \int_0^{r_\alpha} \dfrac{dr}{1-r^2/4}\\
&=2\ln \frac{1}{\tan(\delta_\alpha/2) }\\
 &=2 \alpha  \ln \dfrac{1}{\tan(\delta/2)}\\
 &\approx 2 \alpha  \ln \dfrac{2}{\delta}.
 \end{aligned}
\end{equation}
To study a fixed-length state with prescribed length $L_*$, we then use \eqref{eq:alphamu} and \eqref{eq:Clength} to determine the required tension $\mu$ of the cosmic string.  Below, from \eqref{eq:Clength} we keep only the leading order term at small $\delta$ .

We may also use the above results to compute the (cutoff) length of a RT surface defined by an interval $R$ of any angular size $2\lambda$ .  The point here is that the isometries of global AdS$_3$ act as conformal transformations on the boundary $S^2$ and can be used to map the interval $[0,\pi]$ to the interval $[0,2\lambda]$. Such isometries are easy to describe by embedding Euclidean AdS$_3$ into four-dimensional Minkowski space. In our coordinates this embedding takes the form:
\begin{eqnarray}
\begin{aligned}
T&=\frac{4+r^2}{4-r^2},\\
X&=\frac{4r}{4-r^2}\sin \tilde \theta\sin\phi,\\
Y&=\frac{4r}{4-r^2}\cos \tilde \theta,\\
Z&=\frac{4r}{4-r^2}\sin \tilde \theta\cos\phi .
\end{aligned}
\end{eqnarray}
While the above embedding holds only for the case $\alpha=1$ (where $\tilde \theta=\theta$), we have chosen to write the embedding in terms of $\tilde \theta$ as we will eventually apply the boundary conformal transformation to cases with general $\alpha$ using the round conformal frame.

It will be convenient to take the new interval $R$ to also lie along the boundary great circle defined by $\phi=0$ and $\phi=\pi$.  Note that such intervals all lie at $X=0$ in the embedding coordinates, and that they are thus invariant under the $\mathbb{Z}_2$ isometry $(T,X,Y,Z) \rightarrow (T,-X,Y,Z)$.  We refer to this isometry as reflection in $X$.

Note that boosts in the $Z,T$ plane preserve this $\mathbb{Z}_2$ symmetry while acting non-trivially on the desired boundary great circle.  In particular, a boost in the negative $Z$ direction with rapidity $\eta$ acts on this circle as $\sin \tilde \theta \rightarrow \frac{\sin \tilde \theta - \eta}{1 -\eta \sin \tilde \theta}$.  So to map the angular interval $\tilde \theta \in [\pi/2 -\lambda, \pi/2+ \lambda]$ at $\phi=0$ to the interval $\tilde \theta \in [0, \pi]$ at $\phi=0$ we need only choose $\eta = \sin \lambda$.

Such a boost also acts on our cutoff, taking a cutoff $\delta$ associated with  $\tilde \theta \in [\pi/2 -\lambda, \pi/2+ \lambda]$  to a new cutoff associated with $\tilde \theta \in [0, \pi]$  given by
\begin{equation}
\label{eq:deltab}
\begin{aligned}
\delta_b & =\frac{1}{2}\left(\sin^{-1}\frac{\sin(\frac{\pi}{2}-\lambda+\delta)-\sin(\frac{\pi}{2}-\lambda) }{1-\sin (\frac{\pi}{2}-\lambda+\delta)\sin(\frac{\pi}{2}-\lambda)}-\sin^{-1}\frac{\sin(\frac{\pi}{2}-\lambda-\delta)-\sin(\frac{\pi}{2}-\lambda) }{1-\sin (\frac{\pi}{2}-\lambda-\delta)\sin(\frac{\pi}{2}-\lambda)}\right)\\
& =\frac{\delta}{\sin \lambda} +O(\delta^2).
\end{aligned}
\end{equation}

Applying the associated boundary conformal transformation to the general case $\alpha \neq 1$, we then find that boundary intervals of angular size $\lambda$ are associated with bulk cosmic strings of length given by \eqref{eq:Clength} with $\delta$ replaced by \eqref{eq:deltab} to yield
\begin{equation}
\label{eq:SingIntresult}
L=2\alpha \ln \frac{2\sin\lambda}{\delta} = : \alpha L_0(\lambda).
\end{equation}
where $L_0(\lambda)$ is the cutoff length of this same geodesic when there is no cosmic string.

In section \ref{Sec:Examples} below, we will also find it useful to allow different cutoffs $\delta_L$ and $\delta_R$ at the two ends of the cosmic string. Generalizing the above arguments then yields
\begin{equation}
\label{eq:LRdelta}
L=\alpha \ln \frac{4\sin^2 \lambda}{\delta_L\delta_R},
\end{equation}
where $\alpha$ again describes the defect on this string.

It is now straightforward to compute the Euclidean action $I$ of our solutions as a function of $\lambda$, $\delta$, and $L=L_*= \alpha L_0$.  Since this computation is somewhat of an aside from the main thrust of this work we have relegated the details to appendix \ref{app}.  Up to an $\alpha$-independent constant (which depends on the choice of boundary conformal frame, and thus in a fixed frame may depend on $\delta$ and $\lambda$), the action can be written in terms of just $\alpha$ and $L_0$:
\begin{equation}
\label{eq:oneint}
I=\frac{\alpha (\alpha-2) L_0}{8G}.
\end{equation}

Since there is no RT phase transition for single intervals, we can use the results of \cite{Harlow:2016vwg} to write the density matrix of the dual CFT on our interval in the block-diagonal form
\begin{equation}
\label{eq:rho1}
\rho=\oplus_{\alpha}P(\alpha) \rho_\alpha,
\end{equation}
where $P(\alpha)$ is the probability for the RT surface to have length $\alpha L_0$.  As explained in section \ref{Sec:Review}, in the semiclassical approximation this probability is
\begin{equation}
P(\alpha)=N\exp(-I)=\sqrt{\dfrac{L_0}{8\pi G}}\exp\left(-\dfrac{ (\alpha-1)^2 L_0}{8G}\right),
\end{equation}
where in the last step we have computed the appropriate normalization coefficient $N$ so that\footnote{In fact, we have used the value of $N$ for which
$1= \int_{-\infty}^\infty  P(\alpha) d\alpha = \int_{0}^\infty  P(\alpha) d\alpha  + O\left(e^{-\dfrac{ L_0}{8G}}\right)$.  The associated error is negligible in the semiclassical limit.} $\int_{0}^\infty  P(\alpha) d\alpha = 1$.  Furthermore, in this approximation reference \cite{Dong:2018seb} finds each $\rho_\alpha$ to be a maximally mixed state in a subspace whose dimension agrees with the RT entropy $L_0/4G$.  Thus we may write
\begin{equation}
\label{eq:alpharho}
\rho_\alpha=e^{-\frac{\alpha L_0}{4G}} \mathbb{I}_{\exp\left({\frac{\alpha L_0}{4G}}\right)},
\end{equation}
where $ \mathbb{I}_{\exp\left(S\right)}$ is the identity matrix in a Hilbert space of dimension $e^S$.

The physics of the result \eqref{eq:alpharho} is most easily seen as follows.  Let us focus on the case $\lambda = \pi/2$ for simplicity, and let us then identify the cutoff surface near $\tilde \theta = 0$ with the one near $\tilde \theta = \pi$.  Except for the conical singularity, the resulting spacetime is a Euclidean BTZ black hole with horizon length $L_* = \alpha L$ and energy $E_{BTZ}= \frac{r_+^2}{8G} = \frac{L^2}{32\pi^2 G}$ as defined in the standard BTZ conformal frame.  If we treat $\alpha$ (and thus $E$) as a discrete index, we may then write the density matrix \eqref{eq:rho1} as
\begin{equation}
\label{eq:rho1E}
\rho=N_E \left( \oplus_{E} e^{-\beta E}  \mathbb{I}_{\exp\left(S_{BTZ}(E)\right)} \right),
\end{equation}
where $\beta_{BTZ} = 4\pi^2/L_0$ and $S_{BTZ}(E) = \sqrt{2\pi^2 E/G}$ is the entropy of a BTZ black hole with energy $E$.  The normalization coefficient $N_E$ is
$N_E = \sqrt{\dfrac{L_0}{8\pi G}} e^{- L_0/8G}.$    In other words, the density matrix coincides with a canonical ensemble of BTZ microstates at inverse temperature $\beta$.  This is precisely what one expects from the general discussion of fixed-area states in section 5 of \cite{Dong:2018seb}.

Using the above results, it is of course straightforward to compute R\'enyi entropies.  We find
\begin{equation}
\begin{aligned}
\Tr \rho^n &= \int  P(\alpha)^n e^{-n\frac{\alpha L_0}{4G}} e^{\frac{\alpha L_0}{4G}} d\alpha \\
&=\sqrt{\dfrac{8\pi G}{n L_0}}\left(\dfrac{L_0}{8\pi G}\right)^{n/2}\exp\left( -\dfrac{L_0}{8G} (n-\dfrac{1}{n})\right),
\end{aligned}
\end{equation}
and thus
\begin{equation}
\begin{aligned}
S_n &:=\dfrac{1}{1-n} \ln \Tr \rho^n\\
&=\dfrac{L_0}{8G}\left(1+\dfrac{1}{n}\right)+O(\ln(G))\\
&=\dfrac{c}{6}\left(1+\dfrac{1}{n}\right)\ln \frac{2\sin \lambda}{\delta}+O(\ln(c)),
\end{aligned}
\end{equation}
where we used the Brown-Henneaux relation $c=\frac{3\ell}{2G}$ \cite{Brown:1986nw}, with $\ell=1$. Of course, this precisely matches the well-known results of \cite{Calabrese:2004eu, Calabrese:2009qy} for the dual CFT.

\section{Examples}
\label{Sec:Examples}

We now we consider several examples of the general framework discussed above.  The first two cases concern AdS$_3$ and its BTZ quotients.  In those cases we compute the covariance matrix \eqref{eq:cov} by treating the conical defect as a small perturbation, working to linear order in the (Euclidean) tension $\mu$ of the associated (spacelike) cosmic strings.  As a result, the effect of multiple such cosmic strings satisfy linear superposition, and results for general configurations of strings can be computed from the one-interval results of section \ref{Sec:SingleInterval}.  In practice,  instead of the fixed-area action $I$, we find it convenient to study the action $I_{def} = I_{defect}$ for fixed tensions $\mu_1,\mu_2$ of the cosmic strings along the two RT surfaces.  However, the two are related by a Legendre transform $I = I_{def} - \mu_1 A_1 - \mu_2 A_2$ (see e.g. \cite{Dong:2018seb,Dong:2019piw}).  As a result, the matrix $\frac{\partial^2 I}{\partial A_i \partial A_j}$ is the inverse of $\frac{\partial^2 I}{\partial \mu_i \partial \mu_j}$ and we have
\begin{equation}
\label{eq:mucov}
C_{ij} = \frac{\partial^2 I_{def}}{\partial \mu_i \partial \mu_j} = - \frac{\partial}{\partial \mu_i} \langle A_j \rangle_{\mu_1,\mu_2} \Big|_{\mu_1 = \mu_2 = 0}=  - \frac{\partial}{\partial \mu_{j}} \langle A_i  \rangle_{\mu_1,\mu_2} \Big|_{\mu_1 = \mu_2 = 0},
\end{equation}
where $\langle A_i \rangle_{\mu_1,\mu_2}$ is the most likely value of $A_i$ in the presence of cosmic strings with tensions $\mu_1,\mu_2$.  Here we have used the standard Legendre transform relation $\frac{\partial I_{def}}{\partial \mu_i} = - \langle A_i \rangle_{\mu_1,\mu_2}$.

\subsection{Example 1: Two intervals in the $AdS_3$ vacuum}
\label{Sec:TwoIntervals}

Our first example concerns the Euclidean global $AdS_3$ vacuum as in section \ref{Sec:SingleInterval}.  However, we now take the boundary region $R$ to be given by a pair of non-overlapping intervals on the great circle of the boundary $S^2$ associated with $\phi =0$ and $\phi = \pi$.   For simplicity, we choose the two intervals to be related by a $\pi$ rotation.  In particular, they are each of the same angular size $2\lambda < \pi$.  We take both to be given by $\theta \in [\pi/2-\lambda, \pi/2+\lambda]$ and to respectively lie at $\phi=0$ and $\phi = \pi$.

As is well known, there are two locally-minimal surfaces that satisfy the required boundary conditions.  While both are homologous to the pair of boundary intervals $R$, only one of them is homotopic to $R$.  For reasons that will shortly become clear, we denote this homotopic surface by $\gamma_d=\gamma_{diagonal}$ while the other will be denoted $\gamma_{o}=\gamma_{off-diagonal}$. Since the RT surfaces are one-dimensional, we will again use the terms length and area interchangeably as in section \ref{Sec:SingleInterval}. In particular, the total lengths of the above RT surfaces are $L_d$ and $L_o$.

\begin{figure}
\centering
\begin{tikzpicture}
\draw (0,0) circle [radius=3];
\draw [red] (2.3,1.93) to [out=220,in=140] (2.3,-1.93);
\draw [red] (-2.3,1.93) to [out=-40,in=40] (-2.3,-1.93);
\draw [blue] (-2.3,1.93) to [out=-40,in=220] (2.3,1.93);
\draw [blue] (-2.3,-1.93) to [out=40,in=140] (2.3,-1.93);
\node at (3.5,0) {$R_2$};
\node at (-3.5,0) {$R_1$};
\node at (-2,0) {$\gamma_{11}$};
\node at (2,0) {$\gamma_{22}$};
\node at (0,1.5) {$\gamma_{12}$};
\node at (0,-1.5) {$\gamma_{21}$};
\end{tikzpicture}
\caption{Two competing extremal surfaces when $R= R_1 \cup R_2$ is a pair of intervals on the boundary of the $AdS_3$ vacuum.  The homotopic RT surface $\gamma_d = \gamma_{11} \cup \gamma_{22}$ and the  non-homotopic RT surface $\gamma_o = \gamma_{12} \cup \gamma_{21}$ are shown respectively in red and blue. The case shown sits precisely at the RT phase transition, where $\gamma_d$ and $\gamma_o$ are related by a $\pi/2$ rotation.}
\label{fig:gammaod}
\end{figure}
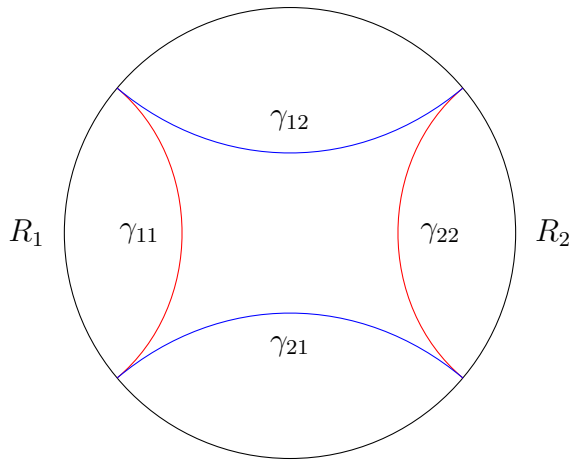

Each of the above RT surfaces is disconnected, and in fact consists of two geodesics.  We label the four relevant geodesics $\gamma_{11}, \gamma_{12}, \gamma_{21}, \gamma_{22}$, with $\gamma_d = \gamma_{11} \cup \gamma_{22}$ and $\gamma_{o} = \gamma_{12} \cup \gamma_{21}$ as shown in figure \ref{fig:gammaod}.  The corresponding lengths are $L_{11}=L_{22}$, and $L_{12}=L_{21}$.    The system undergoes an RT phase transition at $\lambda=\pi/4$, when $\gamma_d$ and $\gamma_o$ are related by a $\pi/2$ rotation. For vanishing cosmic-string tensions the solution is just global Euclidean AdS$_3$ and the lengths of the RT surfaces are
\begin{equation}
\bar{L}_d=2\bar{L}_{11}=2\bar{L}_{22}=4\ln \frac{2\sin\lambda}{\delta}
\end{equation}
\begin{equation}
\bar{L}_o=2\bar{L}_{21}=2\bar{L}_{12}=4\ln \frac{2\cos\lambda}{\delta}.
\end{equation}

Due to the superposition principle mentioned in the introduction to this section, it will be convenient to allow independent cosmic string tensions $\mu_{ij}$ for all $i,j \in \{1,2\}$. To compute \eqref{eq:mucov}, we need only find the response functions $\Delta_{mn}L_{ij}$ that describe how the lengths $L_{ij}$ of the geodesics in figure \ref{fig:gammaod} change at linear order under the addition of the sources $\mu_{mn}$.  Of the 16 response functions $\Delta_{mn}L_{ij}$,  the four terms $\Delta_{ij}L_{ij}$ where we study the change in length $L_{ij}$ along the same defect (with tension $\mu_{ij}$) are just the linearization of the single-interval result \eqref{eq:SingIntresult} from section \ref{Sec:SingleInterval}.  Furthermore, the 8 terms $\Delta_{mn}L_{ij}$ where $(m,n,i,j)$ are permutations of $(1,1,1,2)$ and $(2,2,2,1)$ (i.e., where 3 of the 4 indices $m,n,i,j$ coincide but the last is different) are all related to each other by symmetry (and perhaps interchanging $\lambda \rightarrow \pi/2 -\lambda$).  Finally, the last 4 terms $\Delta_{ij}L_{\bar i \bar j}$ (with $\bar i \neq i$ and $\bar j \neq j$) involve diametrically opposite geodesics.  As representatives of these 3 classes of terms, we will compute $\Delta_{11}L_{11}$, $\Delta_{11} L_{12}$, and $\Delta_{11}L_{22}$.

Let us begin by computing $\Delta_{11}L_{12}$, the first-order change in the length $L_{12}$ due to the source $\mu_{11}$.  As in section \ref{Sec:SingleInterval}, this is straightforward if we act with an AdS isometry to move the boundary-anchors of $\gamma_{11}$ to the poles $\theta =0, \pi$, so that $\gamma_{11}$ runs along the $\phi$-axis; see figure \ref{fig:2intboost} below.   After this transformation, the two anchors of $\gamma_{12}$ lie at the pole $\theta =0$ and at $\theta = \vartheta$ with
\begin{equation}
\label{eq:vartheta}
\vartheta =\sin^{-1} \frac{2\cos{\lambda}}{1+\cos^2{\lambda}}.
\end{equation}

\begin{figure}
\centering
\begin{tikzpicture}
\draw (0,0) circle [radius=3];
\draw [red] (2.8,1.02) to [out=200,in=160] (2.8,-1.02);
\draw [red] (0,3) to (0,-3);
\draw [blue] (0,3) to [out=-90,in=200] (2.8,1.02);
\draw [blue] (0,-3) to [out=90,in=160] (2.8,-1.02);
\node at (-0.5,0) {$\gamma_{11}$};
\node at (1.8,0) {$\gamma_{22}$};
\node at (1.5,1.6) {$\gamma_{12}$};
\node at (1.5,-1.6) {$\gamma_{21}$};
\end{tikzpicture}
\caption{The same geodesics as in figure \ref{fig:gammaod} after applying an AdS$_3$ isometry to move the boundary-anchors $\gamma_{11}$ to the poles $\theta=0,\pi$.  This configuration allows us to compute changes in length by applying results from section \ref{Sec:SingleInterval}.}
\label{fig:2intboost}
\end{figure}
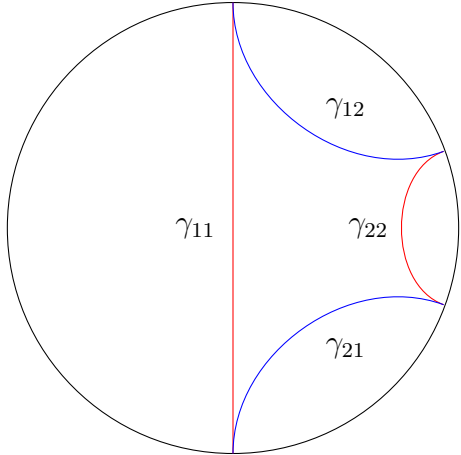

In the resulting (round) conformal frame, the cutoffs at the two ends of $L_{12}$ will differ. This occurs because the boundary conformal transformation associated with the above AdS$_3$ isometry fails to preserve the original symmetry between the endpoints.  In the new conformal frame the cutoffs are given by
\begin{equation}
\delta_L \approx\frac{\delta}{\sin{\lambda}}
\end{equation}
\begin{equation}
\label{eq:2intdeltaR}
\begin{aligned}
\delta_R &=\frac{1}{2}\left(\sin^{-1}\frac{\sin(\frac{\pi}{2}-\lambda+\delta)+\sin(\frac{\pi}{2}-\lambda)}{1+\sin(\frac{\pi}{2}-\lambda+\delta)\sin(\frac{\pi}{2}-\lambda)}-\sin^{-1}\frac{\sin(\frac{\pi}{2}-\lambda-\delta)+\sin(\frac{\pi}{2}-\lambda)}{1+\sin(\frac{\pi}{2}-\lambda-\delta)\sin(\frac{\pi}{2}-\lambda)}\right) \\
&\approx\frac{2\sin{\lambda}}{3+\cos{2\lambda}}\delta,
\end{aligned}
\end{equation}
with $\delta_L$ being the cutoff at left end in figure \ref{fig:2intboost}, where $\gamma_{12}$ meets $\gamma_{11}$.

Now, the length of a geodesic in the position of $\gamma_{12}$ with cutoffs $\delta_L,\delta_R$ in pure AdS$_3$ was studied in section \ref{Sec:SingleInterval}.  It was found there to be given by \eqref{eq:LRdelta}, where one should insert the value $\alpha=1$ since there is no defect on $\gamma_{12}$. And since the local metric near $\gamma_{12}$ in coordinates $(r, \theta, \phi)$ does not change when we insert a string of tension $\mu_{11} = \frac{1 - \alpha_{11}}{4G}$ on $L_{11}$, the length $L_{12}$ in the presence of this defect can again be obtained from \eqref{eq:LRdelta} with $\alpha=1$.  However, after the addition of the defect the coordinates $(r, \theta, \phi)$ are associated with the conical conformal frame on the boundary.  As a result, we must insert into \eqref{eq:LRdelta}  the $\theta$-locations $0,\hat \vartheta$ of the $\gamma_{12}$ anchors and the cutoffs $\hat \delta_L,\hat \delta_R$ as described in the conical conformal frame.  Using again the conformal transformation \eqref{eq:thetas}, we find
\begin{equation}
\hat{\vartheta}=2 \tan^{-1} \left(\tan^{\alpha_{11}} \frac{\vartheta}{2}\right),
\end{equation}
\begin{equation}
\hat{\delta}_L=2 \tan^{-1} \left(\tan^{\alpha_{11}} \frac{\delta_L}{2}\right) \approx 2\left(\frac{\delta_L}{2}\right)^{\alpha_{11}},
\end{equation}
and
\begin{equation}
\begin{aligned}
\hat{\delta}_R &=\tan^{-1}\left(\tan^{\alpha_{11}}\frac{\vartheta+\delta_R}{2}\right) - \tan^{-1}\left(\tan^{\alpha_{11}}\frac{\vartheta-\delta_R}{2}\right) \\
&\approx \frac{{\alpha_{11}}(3+\cos{2\lambda})\sin^{{\alpha_{11}}-1}{\lambda}}{2(1+\cos^{2{\alpha_{11}}}{\lambda})}\delta_R.
\end{aligned}
\end{equation}
Here the symbol $\approx$ indicates that we have dropped higher order terms in the original cutoff $\delta$.
The first-order change in length is thus
\begin{equation}
\begin{aligned}
\Delta_{11}L_{12}&=\ln\frac{4\sin^2 \frac{\hat{\vartheta}}{2}}{\hat{\delta}_L\hat{\delta}_R}-\ln\frac{4\sin^2\frac{\vartheta}{2}}{\delta_L\delta_R}\\
&\approx4\mu_{11} G\left(1-\ln\frac{\sin2\lambda}{\delta}\right).
\end{aligned}
\end{equation}

We now address the diametrically opposite case.  In particular, we compute the change $\Delta_{11} L_{22}$ in $L_{22}$ when we add tension $\mu_{11}$ on $\gamma_{11}$.  As above, we apply an AdS$_3$ isometry to move the anchors of $\gamma_{11}$ to the poles as shown in figure \ref{fig:2intboost}. Since this figure is symmetric under exchange of the two ends of $\gamma_{22}$, and since the left end of $\gamma_{22}$ coincides with the right end of $\gamma_{12}$, after the transformation the cutoff at either end of $\gamma_{22}$ becomes $\delta'=\delta_R$ as given by \eqref{eq:2intdeltaR} and the angular size of $\gamma_{22}$ becomes $2\lambda'=2(\frac{\pi}{2}-\vartheta)$ in terms of \eqref{eq:vartheta}.

Once again, we wish to hold fixed the locations and cutoffs in the round conformal frame when we insert the cosmic string on $\gamma_{11}$.  And again we wish to apply formulae from section \ref{Sec:SingleInterval} that apply in the conical-frame coordinates $r,\theta, \phi$. We will thus need the associated conical frame cutoff $\hat \delta=\hat{\delta}_R$ and angular size $2\hat \lambda=2(\frac{\pi}{2}-\hat{\vartheta})$. The length change is thus
\begin{equation}
\begin{aligned}
\Delta_{11} L_{22}&=2\ln \frac{2\sin\hat{\lambda}}{\hat {\delta}}-2 \ln\frac{2\sin\lambda}{\delta'}\\
&\approx 8\mu_{11} G\left(1+\left(\frac{2}{\sin^2\lambda}-1\right)\ln\cos \lambda \right).
\end{aligned}
\end{equation}

To complete our study of the 3 possible classes of changes we need only compute $\Delta_{11}L_{11}$. From \eqref{eq:SingIntresult} we immediately find
\begin{equation}
\Delta_{11}L_{11}=-8\mu_{11} G \ln\frac{2\sin\lambda}{\delta}.
\end{equation}

We are now ready to assemble the above results into complete expressions for the first order changes in our lengths.
As described above, our three representatives $\Delta_{11}L_{11}$, $\Delta_{11} L_{12}$, and $\Delta_{11}L_{22}$ can be used to obtain all other $\Delta_{mn}L_{ij}$ by acting with appropriate symmetries and/or replacing $\lambda$ by $\frac{\pi}{2}-\lambda$.  After doing so, we wish to set the tension to be constant along each of $\gamma_d$, $\gamma_o$.  I.e., we impose $\mu_{11}=\mu_{22}=\mu_{d}$, $\mu_{12}=\mu_{21}=\mu_{o}$.  Using the notation $\langle L_d\rangle_{\mu_o,\mu_d}$ for the expectation value of $L_d$ in the presence of sources, we have
\begin{equation}
\langle L_d \rangle_{\mu_o,\mu_d}= 2\langle L_{11} \rangle_{\mu_o,\mu_d}=2\left(\langle {L}_{11}\rangle_{0,0}+\Delta_{11}L_{11}+\Delta_{12}L_{11}+\Delta_{21}L_{11}+\Delta_{22}L_{11}\right) + O(\mu^2).
\end{equation}
Thus we find
\begin{equation}
\label{eq:Ld}
\begin{aligned}
\langle L_d \rangle_{\mu_o,\mu_d}=4\ln \frac{2\sin\lambda}{\delta} &+16\mu_d G(1-\ln\sin{2\lambda} + \frac{2\ln\cos{\lambda}}{\sin^2{\lambda}} + \ln\delta) \\ &+ 16\mu_o G(1-\ln\sin{2\lambda}+ \ln\delta) +O(\mu^2).
\end{aligned}
\end{equation}
The corresponding expression for $L_o$ is obtained from \eqref{eq:Ld} by exchanging $\mu_d$ with $\mu_o$ and replacing $\lambda$ by $\pi/2-\lambda$.  This yields
\begin{equation}
\begin{aligned}
\langle L_o \rangle_{\mu_o,\mu_d}=4\ln \frac{2\cos\lambda}{\delta} &+16\mu_d G(1-\ln\sin{2\lambda}+ \ln\delta)
\\ &+16\mu_o G(1-\ln\sin{2\lambda} + \frac{2\ln\sin{\lambda}}{\cos^2{\lambda}} + \ln\delta) +O(\mu^2).
\end{aligned}
\end{equation}
The two point functions are thus
\begin{equation}
\label{eq:sigd}
\langle L_d^2\rangle_{0,0} -\langle L_d\rangle^2_{0,0}=-\frac{\partial}{\partial \mu_d} \langle L_d\rangle_{\mu_o,\mu_d}\Big|_{\mu_o=0,\mu_d=0} =-16G(1-\ln\sin{2\lambda} + \frac{2\ln\cos{\lambda}}{\sin^2{\lambda}} + \ln\delta),
\end{equation}
\begin{equation}
\label{eq:sigo}
\langle L_o^2\rangle_{0,0} -\langle L_o\rangle^2_{0,0}=-\frac{\partial}{\partial \mu_o}\langle L_o\rangle_{\mu_o,\mu_d}\Big|_{\mu_o=0,\mu_d=0}=-16G(1-\ln\sin{2\lambda} + \frac{2\ln\sin{\lambda}}{\cos^2{\lambda}} + \ln\delta),
\end{equation}
and
\begin{equation}
\label{eq:sigc}
\langle L_d L_o\rangle_{0,0} -\langle L_d\rangle \langle L_o\rangle_{0,0}=-\frac{\partial}{\partial \mu_d}\langle_{\mu_o,\mu_d} L_o\rangle\Big|_{\mu_o=0,\mu_d=0}=-16G(1-\ln\sin{2\lambda}+ \ln\delta).
\end{equation}
Combining these to find the variance of $(L_1-L_2)$ yields
\begin{equation}
\begin{aligned}
4\sigma^2_{-}&=\langle(L_d-L_o)^2\rangle_{0,0}-\langle L_d-L_o\rangle^2_{0,0}\\
&=\left(\langle L_d^2\rangle -\langle L_d\rangle^2_{0,0}\right)+\left(\langle L_o^2\rangle_{0,0} -\langle L_o\rangle^2_{0,0}\right)-2\left(\langle L_d L_o\rangle_{0,0} -\langle L_d\rangle_{0,0} \langle L_o\rangle_{0,0}\right)\\
&=-32G \left( \frac{\ln \sin\lambda}{\cos^2\lambda}+\frac{\ln \cos\lambda}{\sin^2\lambda}\right)
\end{aligned}
\end{equation}
Note that $\sigma_-^2$ is positive as required since $\cos \lambda$ and $\sin \lambda$ are less than or equal to one.
From \eqref{eq:entcor}, the $O(G^{-1/2})$ correction to the entropy at the transition is thus
\begin{equation}
\label{eq:2intScor}
\Delta_{-1/2} S= \sqrt{\frac{\tilde  \sigma^2_-}{8\pi G}} = \frac{\sigma_-}{\sqrt{8\pi}G} = 2 \sqrt{\frac{\ln 2}{2\pi G}}.
\end{equation}

The most interesting feature of \eqref{eq:2intScor} is that it is independent of the cutoff $\delta$.  This was a direct result of the fact that, while the cutoff appeared in each of \eqref{eq:sigd}, \eqref{eq:sigo}, and \eqref{eq:sigc}, it cancelled in the computation of $\tilde \sigma_-$.
A related observation is that $\sigma^2_{-}$ takes on its minimal value $16 G\ln 2$ at the phase transition point $\lambda = \pi/4$, though it diverges in the degenerate limits $\lambda\rightarrow 0$ or $\lambda \rightarrow \pi/2$.

Such results are in fact very natural.  Since $\gamma_o$ and $\gamma_d$ have the same boundary anchors, the two curves will largely coincide near infinity.  Contributions to the length of these curves from the asymptotic region will thus be highly correlated and will tend to cancel in computations of $L_d - L_o$.  The results above show that the divergent parts of the fluctuations cancel entirely.  Thus $\sigma_-$ is determined by the regions of $\gamma_d$ and $\gamma_o$ that are widely separated.  Since the length of such regions diverges in the limits $\lambda\rightarrow 0$ or $\lambda \rightarrow \pi/2$ (where one curve or the other degenerates), it is no surprise that $\sigma_-$ diverges in those limits as well.  Further discussion of divergences in RT-area fluctuations will be provided in section \ref{Sec:Disc}.

\subsection{Example 2: BTZ black hole}\label{Sec:BTZ}
Our next example is a generic pure microstate of a one-sided non-rotating BTZ black hole.  The RT phase transition in this context was previously studied in
e.g. \cite{Hubeny:2013gta,Asplund:2014coa,Bao:2017guc}.
Since the bulk spacetime has dimension 3, our RT surfaces will again be spacelike geodesics.

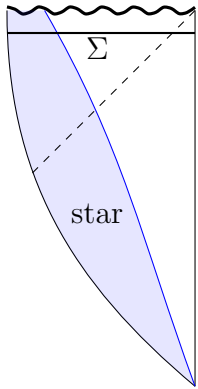
\begin{figure}
\centering
\begin{tikzpicture}
\draw (3,0) -- (3,5);
\draw [domain=0.5:3,smooth,very thick,variable=\x] plot ({\x},{5+0.05*sin((15*\x) r)});
\draw [name path=A] (3,0) to [out=140,in=-90] (0.5,5);
\draw [name path=B,blue] (3,0) to [out=110,in=-60] (1,5);
\tikzfillbetween[of=A and B]{blue, opacity=0.1};
\draw [dashed] (3,5) to (1-0.2,3-0.2);
\node at (1.7,2.3) {star};
\draw [thick] (3,4.7) to (0.5,4.7);
\node at (1.7,4.5) {$\Sigma$};
\end{tikzpicture}
\caption{The Penrose diagram of a pure-state geometry in which matter (a `star') collapses to become a BTZ black hole. On the surface $\Sigma$ the geometry displays a long `throat' region where the metric is {\it very} well-approximated by exact BTZ. On $\Sigma$, all details of the original star that collapsed to form the black hole are hidden at the bottom of this throat.}
\label{fig:collapse}
\end{figure}

In the region outside the horizon, the corresponding bulk geometry should well-described by the BTZ metric
\begin{equation}
\label{BTZmetric}
ds^2=-(r^2_{BTZ}-r_+^2)dt^2+\frac{dr^2_{BTZ}}{r^2_{BTZ}-r_+^2}+r^2_{BTZ}d\varphi^2,
\end{equation}
where $r_+$ is the horizon radius and the black hole has total energy $E = \frac{r_+^2}{8G}$ as in section \ref{Sec:SingleInterval}.
Inside the horizon the geometry may reflect the details of the microstate.  But as shown in figure \ref{fig:collapse}, any classical interior solution will evolve to have the same long throat at late times, with any microstate-dependence hidden at the bottom of the throat.  One thus expects such a long throat to be common to generic microstates.  Noting that this throat also appears in the two-sided eternal BTZ black hole, and that in the eternal BTZ case spacelike geodesics starting and ending in the same boundary region never pass behind the horizon (see e.g. \cite{Morrison:2012iz}), it follows that all relevant geodesics will lie in the exterior region.  We thus restrict attention to the geometry described by \eqref{BTZmetric}.

By analytic continuation $t \rightarrow i\tau$, the associated Euclidean solution will also contain a region described by the metric
\begin{equation}
\label{eq:EBTZmetric}
ds^2=(r^2_{BTZ}-r_+^2)d\tau^2+\frac{dr^2_{BTZ}}{r^2_{BTZ}-r_+^2}+r_{BTZ}^2d\varphi^2,
\end{equation}
though this will not cover the entire spacetime.  In particular, the metric \eqref{eq:EBTZmetric} will generally hold only in some range $\tau \in (-\tau_-,\tau_+)$ where $\tau_-$ and $\tau_+$ are {\it not} to be identified\footnote{See e.g. \cite{Krasnov:2000zq,Krasnov:2003ye,Skenderis:2009ju} for discussions of particular such Euclidean geometries.}.
We will consider single intervals $R$ in the boundary at $t=0 = \tau$, for which the RT surfaces will also lie in the bulk surface $t = \tau=0$ that appears in both the Lorentzian and Euclidean sections.

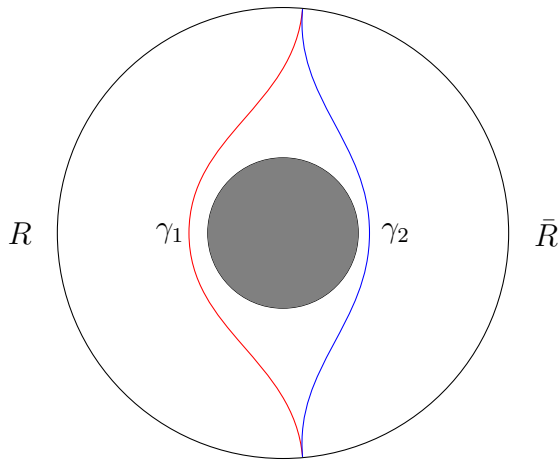
\begin{figure}
\centering
\begin{tikzpicture}
\draw (0,0) circle [radius=3];
\draw (0,0) circle [radius=1];
\fill [color=gray] (0,0) circle [radius=1];
\draw [red] (0.26,2.98) to [out=265,in=90] (-1.25,0) to [out=-90,in=95] (0.26,-2.98)  ;
\draw [blue] (0.26,2.98) to [out=265,in=90] (1.15,0) to [out=-90,in=95] (0.26,-2.98)  ;
\node at (-3.5,0) {$R$};
\node at (3.5,0) {$\bar{R}$};
\node at (-1.5,0) {$\gamma_1$};
\node at (1.5,0) {$\gamma_2$};
\end{tikzpicture}
\caption{Two competing candidate RT surfaces in the BTZ black hole spacetime.}
\label{RTinBTZ}
\end{figure}

As shown in figure \ref{RTinBTZ}, we take $\gamma_1$ to be the minimal curve in the $t=\tau=0$ surface that is homotopic to $R$, and $\gamma_2$ to be the corresponding minimal curve homotopic to $\bar R$. Note that both $\gamma_1$ and $\gamma_2$ are homolologous to both intervals $R, \bar R$. Denoting the angular sizes of $R, \bar R$  respectively by $\pi+\eta$ and $\pi-\eta$, symmetry dictates that there will be an RT phase transition at $\eta =0$.  As in section \ref{Sec:TwoIntervals}, we will insert cosmic strings on $\gamma_1, \gamma_2$ and compute the induced changes in their lengths $L_1,L_2$.  And just as in sections \ref{Sec:SingleInterval} and \ref{Sec:TwoIntervals}, we will again use a UV regulator defined by a scale $\delta$ on the boundary in the conformal frame where the boundary metric is $d\tau^2 + d\varphi^2$.

To compute the desired response functions  $\frac{\partial}{\partial \mu_i}\langle L_j \rangle_{\mu_1,\mu_2}$, we must gain control not only over the original BTZ metric \eqref{eq:EBTZmetric}, but also over solutions deformed by the addition of cosmic strings.
If we were to strictly confine our analysis to the region $-\tau_- < \tau < \tau$ where the metric \eqref{eq:EBTZmetric} applies, this would require a choice of boundary condition at $\tau = \tau_\pm$.  We do not wish to rely on an ad hoc such choice.  But as noted above, in the relevant region of the geometry one expects our spacetime to agree precisely with the eternal {\it two}-sided BTZ black hole.  We will therefore assume that this remains true after the addition of at least weak-tension cosmic strings.  As a result, we simply compute $\frac{\partial}{\partial \mu_i}\langle L_j \rangle_{\mu_1,\mu_2}$ in the full Euclidean BTZ geometry given by \eqref{eq:EBTZmetric} and by taking $\tau$ to have the appropriate period $\beta_{BTZ} = \frac{2\pi}{r_+}$.

As in well-known, the BTZ geometry is a quotient of global AdS$_3$ \cite{Banados:1992gq}.  Lifting the geodesics $\gamma_1,\gamma_2$ to the AdS$_{3}$ cover will allow us to directly apply our previous results from section \ref{Sec:SingleInterval}.  In practice, this lift is accomplished by simply ignoring the fact that $\phi$ is  periodic identified in the BTZ metric\footnote{Since the $\tau$-circle is contractible in BTZ, the coordinate $\tau$ remains periodic with period $\beta_{BTZ}$ in the AdS$_3$ cover.}.  Taking the anchors of both geodesics $\gamma_1, \gamma_2$ to be at $\varphi = \pm\left(\frac{\pi+\eta}{2}  \right)$, we see that $\gamma_1$ lifts to an infinite set of geodesics $\gamma_{1,k}$ anchored at $\varphi = \pm\left(\frac{\pi+\eta}{2}  \right) + 2\pi k$, while $\gamma_2$ lifts to geodesics $\gamma_{2,k}$ anchored at $\varphi = \left(\frac{\pi+\eta}{2}  \right) + 2\pi k$ and $\varphi = -\left(\frac{\pi+\eta}{2}  \right) + 2\pi (k+1)$.  Note that the geodesics $\gamma_{2, -1}, \gamma_{2,0}$ lie on either side of $\gamma_{1,0}$.

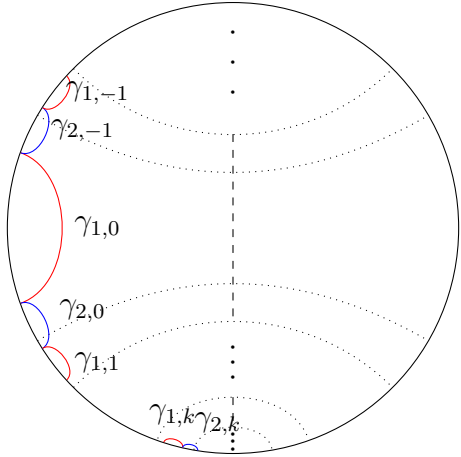
\begin{figure}
\centering
\begin{tikzpicture}
\draw (0,0) circle [radius=3];
\draw [dashed] (0,1.25) to (0,-1.25);
%\draw [dashed] (0,2.25) to (0,2.65);
\draw [dashed] (0,-2.25) to (0,-2.65);
\draw [dotted] (-2.6,1.5) to [out=-30,in=180+30] (2.6,1.5);
\draw [dotted] (-2.6,-1.5) to [out=30,in=180-30] (2.6,-1.5);
\draw [dotted] (-2.12,2.12) to [out=-45,in=225] (2.12,2.12);
\draw [dotted] (-2.12,-2.12) to [out=45,in=135] (2.12,-2.12);
%\draw [dotted] (-1,2.8) to [out=-70,in=-110] (1,2.8);
\draw [dotted] (-1,-2.8) to [out=70,in=110] (1,-2.8);
%\draw [dotted] (-0.52,2.95) to [out=-80,in=-100] (0.52,2.95);
\draw [dotted] (-0.52,-2.95) to [out=80,in=100] (0.52,-2.95);

\draw [red] (-2.82,1) to [out=-20,in=20] (-2.82,-1);
\draw [blue] (-2.53,1.6) to [out=-32.5,in=-20]  (-2.82,1);
\draw [blue] (-2.53,-1.6) to [out=32.5,in=20]  (-2.82,-1);
\draw [red] (-2.53,1.6) to [out=-32.5,in=-42.5]  (-2.21,2.02);
\draw [red] (-2.53,-1.6) to [out=32.5,in=42.5]  (-2.21,-2.02);
\draw [red] (-0.92,-2.85) to [out=72,in=77]  (-0.67,-2.92);
\draw [blue] (-0.67,-2.92) to [out=77,in=81]  (-0.47,-2.96);

\node at (0,2.2) {$\cdot$};\node at (0,2.6) {$\cdot$};\node at (0,1.8) {$\cdot$};
\node at (0,-1.8) {$\cdot$};\node at (0,-1.6) {$\cdot$};\node at (0,-2) {$\cdot$};
%\node at (0,2.75) {$\cdot$};\node at (0,2.85) {$\cdot$};\node at (0,2.95) {$\cdot$};
\node at (0,-2.75) {$\cdot$};\node at (0,-2.85) {$\cdot$};\node at (0,-2.95) {$\cdot$};

\node at (-1.8,0) {$\gamma_{1,0}$};
\node at (-2,1.3) {$\gamma_{2,-1}$};
\node at (-2,-1.1) {$\gamma_{2,0}$};
\node at (-1.8,1.8) {$\gamma_{1,-1}$};
\node at (-1.8,-1.8) {$\gamma_{1,1}$};
\node at (-0.8,-2.5) {$\gamma_{1,k}$};
\node at (-0.2,-2.6) {$\gamma_{2,k}$};
\end{tikzpicture}
\caption{RT surfaces in the $AdS_3$ covering space of a BTZ black hole. One RT surface in BTZ corresponds to infinitely many ones in the covering space. The dashed lines denote the horizon. Since we are studying a one-sided black hole, only half of the space (we take it to be the left half) is relevant to us. }
\label{fig:BTZcover}
\end{figure}

The results of section \ref{Sec:SingleInterval} were written in terms of a different set of coordinates on Euclidean AdS$_3$.  Taking the angular coordinate $\tau$ above to be proportional to $\phi$ of section \ref{Sec:SingleInterval}, one may solve for the relation between our $(r_{BTZ}, \varphi)$ and the $(r,\theta)$ of section \ref{Sec:SingleInterval}.  In particular, on the AdS boundary one finds
\begin{equation}
\theta(\varphi)=\tan^{-1} \left(\sinh(r_+\varphi)\right)+\frac{\pi}{2}.
\end{equation}
Thus if $R$ is the interval $-\frac{\pi}{2}-\frac{\eta}{2}<\varphi<\frac{\pi}{2}+\frac{\eta}{2}$ at $\tau=0$, it also corresponds to the infinite set of intervals $\frac{\pi}{2}-\theta_k<\theta<\frac{\pi}{2}+\theta_k$ with
\begin{equation}
\label{eq:BTZthetak}
\theta_k=\tan^{-1} \left(\sinh\left((\frac{\pi}{2}+\frac{\eta}{2} + 2 \pi k)r_+\right)\right).
\end{equation}

For simplicity, we begin by focusing on the case $k=0$.  For this case, a cutoff $\delta$ defined in terms of the angle $\varphi$ at the endpoints of $R$ maps to a cutoff
\begin{equation}
\begin{aligned}
\delta_0 & =\frac{1}{2}\left(\theta(\frac{\pi}{2}+\frac{\eta}{2}+\delta)-\theta(\frac{\pi}{2}+\frac{\eta}{2}-\delta)\right)\\
& \approx \frac{r_+\cosh\left((\frac{\pi}{2}+\frac{\eta}{2})r_+\right)}{1+\sinh\left((\frac{\pi}{2}+\frac{\eta}{2})r_+\right)}\delta
\end{aligned}
\end{equation}
in terms of the angle $\theta$. With vanishing cosmic string tension the length of $\gamma_1$ is thus
\begin{equation}
\label{eq:BTZbL1}
\bar{L}_1:= \langle L_1 \rangle_{\mu_1=0,\mu_2=0}=2\ln \frac{2\sin \theta_0}{\delta_0}=2\ln \frac{2\sinh\left((\frac{\pi}{2}+\frac{\eta}{2})r_+\right) }{r_+\delta}.
\end{equation}
Since interchanging $R$ and $\bar R$ changes the sign of $\eta$, applying this transformation to \eqref{eq:BTZbL1} yields the length of $\gamma_2$:
\begin{equation}
\label{eq:BTZbL2}
\bar{L}_2:= \langle L_1 \rangle_{\mu_1=0,\mu_2=0}=2\ln \frac{2\sinh\left((\frac{\pi}{2}-\frac{\eta}{2})r_+\right) }{r_+\delta}.
\end{equation}

We now compute the first-order changes $\Delta_i L_j$ ($i,j \in \{1,2\}$) in the length change of $\gamma_j$ due to adding a cosmic string with tension $\mu_i$ on $\gamma_i$. In the covering space description, we could compute the change in length of any of the geodesics $\gamma_{j,n}$.  However, the covering space description of inserting a cosmic string of tension $\mu_i$ on $\gamma_i$ is to in fact insert cosmic strings of this same tension $\mu_i$ on {\it each} of the geodesics $\gamma_{i,k}$. At linear order in $\mu_i$ we may compute the effect of each such cosmic string separately and then simply sum the change induced in our given $\gamma_{j,k}$.

However, performing the above sum is equivalent to inserting a cosmic string on a given geodesic (say, $\gamma_{1,0}$ or $\gamma_{2,0}$), computing the first order change in length for each $\gamma_{1,n}$ or each $\gamma_{2,n}$, and again summing the results.  We will find this perspective to be more convenient in making use of our results from section \ref{Sec:SingleInterval}.   We will thus study the first-order changes $\Delta_1 L_{1,k}$, $\Delta_1 L_{2,k}$ in the lengths of $\gamma_{1,k}, \gamma_{2,k}$ associated with putting
a cosmic string on $\gamma_{1,0}$.  We can then later then obtain results for strings on $\gamma_{2,0}$ by changing the sign of $\eta$.

\begin{figure}
\centering
\begin{tikzpicture}
\draw (0,0) circle [radius=3];
\draw [red] (0,3) to (0,-3);
\draw [blue] (0,-3) to [out=90,in=140] (2.3,-1.92);
\draw [red] (2.81,-1) to [out=160,in=170] (2.95,-0.52);
\draw [blue] (2.95,-0.52) to [out=170,in=175] (2.99,-0.26);
\node at (-0.5,0) {$\gamma_{1,0}$};
\node at (1,-1.5) {$\gamma_{2,0}$};
\node at (2.5,-0.3) {$\gamma_{2,k}$};
\node at (2.3,-0.7) {$\gamma_{1,k}$};
\end{tikzpicture}
\caption{The representative geodesics $\gamma_{1,0}$,  $\gamma_{2,0}$, $\gamma_{1,k}$ and $\gamma_{2,k}$ from figure \ref{fig:BTZcover} are shown after applying an AdS$_3$ isometry to move $\gamma_{1,0}$ into the standard position along the $\phi$-axis.}
\label{BTZboost}
\end{figure}
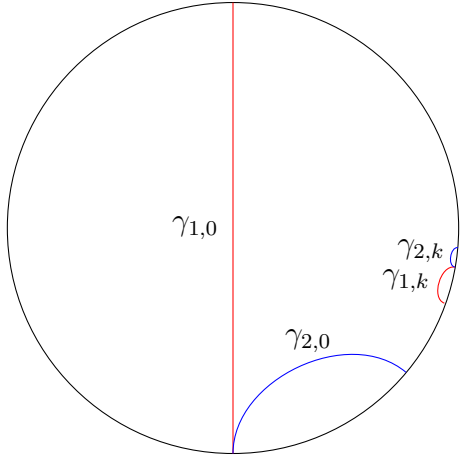

We begin with $\Delta_1 L_{1,k}$. As in section \ref{Sec:SingleInterval}, we apply an AdS$_3$ isometry to move the anchors of $\gamma_{1,0}$ to the $\theta=0$ and $\theta=\pi$ so that $\gamma_{1,0}$ now lies along the $\phi$-axis; see figure \ref{BTZboost}. Before applying this transformation, the angular coordinates $\theta$ of the endpoints of $\gamma_{1,k}$ are
\begin{equation}
\theta_{L/R}^{(k)}=\tan^{-1}\left(\sinh [r_+(\mp \frac{\pi}{2}\mp\frac{\eta}{2}+2\pi k)]\right)+\frac{\pi}{2}
\end{equation}
with $\phi=0$. After the transformation, they become
\begin{equation}
\tilde \theta _{L/R}^{(k)}=\pi-\sin^{-1} \frac{-\sin(\pi-\theta_{L/R}^{(k)})+\cos\theta_1}{1-\sin(\pi-\theta_{L/R}^{(k)})\cos\theta_1},
\end{equation}
with $\phi=\pi$. The cutoffs will again become some $\tilde \delta_L^{(k)}$ and $\tilde \delta_R^{(k)}$.

We now wish to add a cosmic string of tension $\mu_1=\frac{1-\alpha_1}{4G}$ on the $\phi$ axis, holding fixed both the cutoffs and the anchors for the geodesics in the round conformal frame.  But we will also need the values of these parameters in the conical frame, which by \eqref{eq:thetas} are the `hatted' values
\begin{equation}
\hat{\theta}_{L/R}^{(k)}=2\tan^{-1} \left(\tan^{\alpha_1} \frac{\tilde \theta_{L/R}^{(k)}}{2}\right)
\end{equation}
\begin{equation}
\hat{\delta}_{L/R}^{(k)}\approx \frac{2\alpha_1 \cot^{\alpha_1} \frac{\tilde \theta_{L/R}^{(k)}}{2} }{\sin \tilde \theta_{L/R}^{(k)} (1+\cot^{2\alpha_1} \frac{\tilde \theta_{L/R}^{(k)}}{2})}\tilde \delta_{L/R}^{(k)}.
\end{equation}

The above results allow us to read off the change in length by making appropriate use of \eqref{eq:LRdelta}.  This is most straightforward for the case $k=0$ where geodesic $\gamma_{1,0}$ of interest lies on the conical singularity.  That was the setting considered in the derivation of \eqref{eq:SingIntresult}, so we need only recall that our conical parameter is $\alpha_1 = 1- 4\mu_1 G$ and that at $\alpha_1=1$ the length is $\bar L_1$ as given by \eqref{eq:BTZbL1}.  Eq. \eqref{eq:SingIntresult} then gives $L_{1,0} = \alpha_1 \bar L_1$ from which we find
\begin{equation}
\Delta_{1}L_{1,0}= -8\mu_1 G \ln\frac{2\sinh\left((\frac{\pi}{2}+\frac{\eta}{2})r_+\right) }{r_+\delta}.
\end{equation}

In contrast, for $k\neq 1$ we study geodesics $\gamma_{1,k}$ with no conical singularity.  In vacuum AdS$_3$, the length of such geodesics is given by \eqref{eq:LRdelta} with $\alpha =1$ and cutoffs $\delta _R = \tilde \delta_R$ and $\delta_L= \tilde \delta_L$ defined in the round frame.    We wish to hold fixed these round-frame cutoffs when computing $\Delta_{1}L_{1,k}$.  But when we add a conical singularity elsewhere in the spacetime, the length of $\gamma_{1,k}$ is given by \eqref{eq:LRdelta} (still with $\alpha=1$ in that expression) if we insert the cutoffs $\delta_R = \hat \delta_R^{(k)}, \delta_L = \hat \delta_L^{(k)}$ and the angular size $\lambda = \hat \lambda^{(k)} := \frac{\hat \theta_L^{(k)}- \hat \theta_R^{(k)}}{2}$   associated with the {\it conical} frame.  As a result, for $k\neq 0$ we find
\begin{equation}
\label{eq:1k}
\begin{aligned}
\Delta_{1}L_{1,k} &=\ln \frac{4\sin^2 \frac{\hat{\theta}_{L}^{(k)}-\hat{\theta}_{R}^{(k)}}{2}}{\hat{\delta}_{L}^{(k)}\hat{\delta}_{R}^{(k)}}-\ln \frac{4\sin^2\frac{\tilde \theta_{L}^{(k)}-\tilde \theta_{R}^{(k)}}{2}}{\tilde \delta_{L}^{(k)} \tilde \delta_{R}^{(k)}}\\
&=4\mu_1 G\left(2+\frac{\sin\frac{\tilde \theta_{L}^{(k)}+\tilde \theta_{R}^{(k)}}{2}}{\sin\frac{\tilde \theta_{L}^{(k)}-\tilde \theta_{R}^{(k)}}{2}}\ln \frac{\tan \frac{\tilde \theta_{L}^{(k)}}{2}}{\tan \frac{\tilde \theta_{R}^{(k)}}{2}} + O (\mu_1^2)\right)\\
&=4\mu_1 G f(k, \eta) + O(\mu_1^2),
\end{aligned}
\end{equation}
where $f(k,\eta)$ is
\begin{equation}
\begin{aligned}
f(k,\eta) =& 2-\frac{1-2\cosh [2k r_+\pi]+\cosh [r_+(\pi+\eta)]}{2\sinh^2 [(\pi+\eta)\frac{r_+}{2}]}\\
&\ln \frac{\left(\cosh [(\pi+\eta)\frac{r_+}{2}]-\cosh [((4k-1)\pi-\eta)\frac{r_+}{2}]\right)\sinh^2[((2k+1)\pi+\eta)\frac{r_+}{2}]}{\left(\cosh [(\pi+\eta)\frac{r_+}{2}]-\cosh [((4k+1)\pi+\eta)\frac{\r_+}{2}]\right)\sinh^2 [k r_+\pi]}.
\end{aligned}
\end{equation}
Note that $f(k,\eta)$ vanishes exponentially as $k\to \infty$. In particular,  $\lim_{k\to\infty}|\frac{f(k+1,\eta)}{f(k,\eta)}|<1$, so the sum $\sum_{k=1}^\infty f(k,\eta)$ converges.  Furthermore, \eqref{eq:1k} is completely independent of the choice of cutoffs.

For $k\neq -1,0$ the computation of $\Delta_{1}L_{2,k}$ proceeds in precisely the same way. Indeed, it is identical to the computation of
 \eqref{eq:1k}  with $k$ replaced by $k+1/2$ in all expressions and with $\eta$ replaced by $-\eta$ in the expressions for $\theta_{L/R}^{(k)}$ and the associated cutoffs (but with $\eta$ unchanged in the expressions for $\theta_1$ and $\delta_1$).  As a result, we find
\begin{equation}
\Delta_{1}L_{2,k}=4\mu_1 G g(k,\eta) + O(\mu_1^2),
\end{equation}
with
\begin{equation}
\begin{aligned}
g(k,\eta)&=2-\frac{\cosh [r_+\pi]-2\cosh [(1+2k)r_+\pi]+\cosh[r_+\eta]}{\cosh[r_+\pi]-\cosh[r_+\eta]}\\
&\ln \frac{e^{r_+\eta}(e^{2kr_+\pi}-1)(e^{2(1+k)r_+\pi}-1)}{(e^{(1+2k)r_+\pi}-e^{r_+\eta})(e^{r_+(1+2k)\pi+r_+\eta}-1)}.
\end{aligned}
\end{equation}
Note that $g(k,\eta)$ is an even function of $\eta$ as required by the symmetry of figure \ref{fig:BTZcover}.

The remaining cases $\Delta_{1}L_{2,0}$ and $\Delta_{1}L_{2,-1}$ are identical by symmetry; see again figure \ref{fig:BTZcover}.  Let us concentrate on $\Delta_1 L_{2,0}$.  This case differs from the above in that $\gamma_{2,0}$ meets the conical singularity at the boundary.  After using an AdS isometry to place the conical singularity on the $\phi$-axis as usual, the left anchor of $\gamma_{2,0}$ becomes
 $\tilde \theta_L=\pi$ and the right anchor becomes $\tilde \theta_R=\tilde \theta_L^{(1)}$.  The transformed cutoffs are $\tilde \delta_L=\frac{\delta}{\sin \theta_0}$ (with $\theta_0$ again given by \eqref{eq:BTZthetak} with $k=0$) and $\tilde \delta_R=\tilde \delta^{(1)}_L$. As usual, we hold these quantities fixed in the round conformal frame, but we will need to insert the conical frame values into \eqref{eq:SingIntresult}.  After inserting the cosmic string on $\gamma_1^{(0)}$, the conical frame parameters become
 \begin{equation}
 \begin{aligned}
\hat{\theta}_L &=\pi, \ \ \
\hat{\theta}_R =\hat{\theta}^{(1)}_L, \\
\hat{\delta}_L &= 2\left(\frac{\tilde \delta_L}{2}\right)^{\alpha_1} = 2\left(\frac{\delta}{2\sin \theta_0}\right)^{\alpha_1}\\
{\rm and} \ \ \
\hat{\delta}_R &=\hat{\delta}^{(1)}_L.
\end{aligned}
\end{equation}
We thus find
\begin{equation}
\begin{aligned}
\Delta_{1}L_{2,0}
&=4\mu_1 G\left(1-\ln\frac{\cosh r_+ \pi -\cosh r_+ \eta}{r_+ \delta \sinh r_+ \pi}\right).
\end{aligned}
\end{equation}

Combing these results and applying symmetries as needed to obtain changes not directly computed above yields a complete first-order expression for the
length of $\gamma_1$:
\begin{equation}
\begin{aligned}
\langle L_1 \rangle_{\mu_1,\mu_2}=&\bar{L}_1+\Delta_{1}L_{1,0} +2\Delta_{2}L_{1,0}+2\sum_{k=1}^{\infty}\Delta_{1,k}L_{1,0}+2\sum_{k=1}^{\infty}\Delta_{2,k}L_{1,0} +O(\mu^2)\\
=&(2-8\mu_1 G)\ln \frac{2\sinh\left((\frac{\pi}{2}+\frac{\eta}{2})r_+\right) }{r_+\delta} +8\mu_2 G\left(1-\ln\frac{\cosh r_+\pi -\cosh r_+\eta}{r_+\delta \sinh r_+ \pi}\right)\\
&+8\mu_1 G \sum_{k=1}^\infty f(k,\eta)+8\mu_2 G \sum_{k=1}^\infty g(k,\eta) +O(\mu^2)
\end{aligned}
\end{equation}
Since $L_2$ can be obtained from $L_1$ by changing the sign of $\eta$ and exchanging $\mu_1$ and $\mu_2$, we also find
\begin{equation}
\begin{aligned}
\langle L_2 \rangle_{\mu_1,\mu_2}=&(2-8\mu_2 G)\ln \frac{2\sinh\left((\frac{\pi}{2}-\frac{\eta}{2})r_+\right) }{r_+\delta} +8\mu_1 G\left(1-\ln\frac{\cosh r_+\pi -\cosh r_+\eta}{r_+\delta \sinh r_+ \pi}\right)\\
&+8\mu_2 G \sum_{k=1}^\infty f(k,-\eta)+8\mu_1 G \sum_{k=1}^\infty g(k,\eta) +O(\mu^2)
\end{aligned}
\end{equation}

It is now straightforward to compute the desired two-point functions:
\begin{equation} \label{eq:sigma1}
\langle L_1^2\rangle_{0,0} -\langle L_1\rangle ^2_{0.0}=-\frac{\partial}{\partial\mu_1}\langle L_1 \rangle_{\mu_1,\mu_2} \Big|_{\mu_1=\mu_2=0}=8G \left(\ln \frac{2\sinh\left((\frac{\pi}{2}+\frac{\eta}{2})r_+\right) }{r_+\delta}- \sum_{k=1}^\infty f(k,\eta)\right)
\end{equation}
\begin{equation} \label{eq:sigma2}
\langle L_2^2\rangle_{0,0} -\langle L_2\rangle^2_{0,0}=-\frac{\partial}{\partial\mu_2}\langle L_2 \rangle_{\mu_1,\mu_2} \Big|_{\mu_1=\mu_2=0}=8G \left(\ln \frac{2\sinh\left((\frac{\pi}{2}-\frac{\eta}{2})r_+\right) }{r_+\delta}- \sum_{k=1}^\infty f(k,-\eta)\right)
\end{equation}
\begin{equation}
\begin{aligned}
\label{eq:crosscor}
\langle L_1L_2\rangle_{0,0} -\langle L_1\rangle_{0,0} \langle L_2\rangle_{0,0} &=-\frac{\partial}{\partial\mu_1}\langle L_2 \rangle_{\mu_1,\mu_2} \Big|_{\mu_1=\mu_2=0} \\ &=8G\left(\ln\frac{\cosh (r_+\pi) -\cosh (r_+\eta)}{r_+\delta \sinh (r_+ \pi)}-1-\sum_{k=1}^\infty g(k,\eta)\right).
\end{aligned}
\end{equation}
In particular, the variance of $L_- = (L_1-L_2)/2$ is
\begin{equation}
\begin{aligned}
\label{eq:BTZsigma-}
\sigma^2_- =& 2 G\left(\ln \frac{4 \sinh^2(r_+\pi) \sinh \left(\left(\frac{\pi}{2}+\frac{\eta}{2}\right) r_+\right) \sinh \left(\left(\frac{\pi}{2}-\frac{\eta}{2}\right) r_+  \right)}{\left( \cosh(r_+ \pi)-\cosh(r_+ \eta)\right)^2}+2\right.\\
&\left. -\sum_{k=1}^{\infty} \left( f(k, \eta)+f(k,-\eta)-2g(k,\eta)\right)\right).
\end{aligned}
\end{equation}

While the full expressions above are somewhat complicated, one should recall that both $f$ and $g$ fall off exponentially.  As a result, at large $r_+$ one can ignore the sum over images.  In particular, in that limit $\sigma_-^2$ is given by just the first line in \eqref{eq:BTZsigma-}.  As in section \ref{Sec:TwoIntervals}, the variance $\sigma_-^2$ is independent of the cutoff $\delta$, though $\delta$ appears linearly in
$\langle L_1^2\rangle_{0,0} -\langle L_1\rangle ^2_{0.0}$, $\langle L_2^2\rangle_{0,0} -\langle L_2\rangle ^2_{0.0}$, and $\langle L_1L_2\rangle_{0,0} -\langle L_1\rangle_{0,0} \langle L_2\rangle_{0,0}$.

\subsection{Agreement with ETH}
\label{Sec:Matching}

As described in section \ref{Sec:BTZ}, we may think of the analysis performed there as applying to a generic microstate of the BTZ black hole with some given energy $E$.   From the perspective of the dual CFT this is just a generic state with the given energy.   Furthermore, as noted in the introduction, when the volume of the CFT becomes large this reduces to the setting analyzed by Murthy and Srednicki \cite{Murthy:2019qvb} using the eigenstate thermolization hypothesis (ETH).  We now confirm that our results coincide with theirs in the desired limit.

In particular, \cite{Murthy:2019qvb} considered a system of total volume $V$ partitioned into two parts with volumes $V_1 +V_2 =V$.  In the limit where $V_1,V_2$ are both large, and ignoring terms that scale with subleading powers of $V$, we may also identify separate energies $E_1,E_2$ and density-of-states functions\footnote{These are the usual thermodynamic entropies defined as the logarithm of the number of states of each subsystem with the given energies.} $S_1(E_1), S_2(E_2)$ for the two parts that satisfy $E \approx E_1 + E_2$ and $S(E_1,E_2) \approx S_1(E_1) + S_2(E_2)$ .  Here $S(E_1,E_2)$ is the logarithm of the total number of states with the given partition of the energy $E$, and we use the symbol $\approx$ to make explicit that we have kept only terms that are extensive in the sense that they proportional to one of the volumes $V_1$ or $V_2$.    As in section \ref{Sec:BTZ}, we take subsystem $1$ to be associated with the boundary interval $R$ and subsystem $2$ to be associated with $\bar R$.

Typical microstates with energy $E$ will have subsystem energies $\bar E_1,\bar E_2$ determined by the constraint $E=\bar E_1+\bar E_2$ and the usual thermodynamic equilibrium condition
\begin{equation}
\frac{1}{T_1} := \frac{dS_1}{dE_1}|_{\bar E_1} = \frac{dS_2}{dE_2}|_{\bar E_2}  =: \frac{1}{T_2},
\end{equation}
which allows us to define a temperature $T=T_1=T_2$.
The analysis of \cite{Murthy:2019qvb} found such states to have entanglement
\begin{equation}
\label{eq:MS}
S_{ent}(E) =
\min\left( S_1(\bar E_1), S_2(\bar E_2) \right) - \sqrt{\frac{2K}{\pi}}\Phi\left(\frac{S_2(E-\bar{E_1})-S_1(\bar{E_1})}{\sqrt{8K}}\right),
\end{equation}
where $\Phi$ is again given by \eqref{eq:Phi} and
\begin{equation}
\label{eq:Kdef}
\frac{1}{K} : = T^2 \left( \frac{d^2 S_1}{dE_1^2}\rvert_{\bar E_1} +  \frac{d^2 S_2}{dE_2^2}|_{\bar E_2}\right).
\end{equation}
Comparing \eqref{eq:MS} with our expression \eqref{eq:entcor}, we see that they agree if
\begin{equation}
S_1(\bar E_1) = \frac{\bar L_1}{4G},  \ \ \ S_2(\bar E_2) = \frac{\bar L_2}{4G},  \ \ \ K = \frac{\tilde \sigma_-^2}{32G}.
\end{equation}

Our main task is thus to identify the functions $S_1(E_1), S_2(E_2)$ for the relevant limit of the BTZ system studied in section \ref{Sec:BTZ}.  Doing so requires an understanding of black hole geometries that have independent energies $E_1, E_2$ in regions $R$ and $\bar R$ at the given time $t=0$ (though energy will flow between these regions under time evolution due to the intrinsic couplings between the two).  In particular, we must allow the energy densities at $t=0$  to differ between $R$ and $\bar R$.

The limit studied by \cite{Murthy:2019qvb} involves taking a large volume.  But since our system is to be thought of as dual to a conformal field theory, any large volume limit is equivalent to the limit of high temperatures (or, perhaps better, the limit of large energy densities) taken with the volume $V$ held fixed.  We may then define the energy $E_1$ of region $R$ by integrating the CFT energy density over $R$, and similarly for the energy $E_2$ of $\bar R$.  To define a good operator in the CFT we should also apply an appropriate smoothing at the boundary $\partial R$ between $R$ and $\bar R$, though this is often not needed if we simply discuss expectation values.  In either case, we find $E \approx E_1 + E_2$ in the desired limit.

To leading order in the limit of large volumes or high temperatures, we can study the thermodynamics of each region $R$ and $\bar R$ by treating the regions as homogeneous independent CFTs.  The density of states of each region is then given by the thermal entropy of the CFT at energy $E_i$ on a space of volume $V_i$; i.e.
\begin{equation}
\label{eq:Cardy}
S_i(E_i) \approx 2\pi \sqrt{\frac{cE_iV_i}{6}},
\end{equation}
where we have used the Cardy approximation appropriate to our high temperature limit and $c = \frac{3}{2G}$ is the CFT central charge since we have set the bulk AdS scale $\ell$ to one\footnote{One can of course also derive this result from the AdS$_3$ bulk.  To do so, one notes that a general solution to Einstein-Hilbert AdS$_3$ gravity is just a BTZ black hole with some choice of conformal frame.  As we are interested in thermal entropies, so that the full CFT can be in a mixed state, one then computes the RT entropies for $R$ and $\bar R$ using surfaces that are {\it homotopic} to $R$, $\bar R$ as a function of the BTZ parameters and this conformal transformation.  Holding the UV cutoff fixed, at leading order in large energy density maximizing the RT areas at fixed energies $E_1,E_2$ will give the desired result.  Indeed, in a general theory of gravity one should expect the generic high energy-density state with energies $E_1,E_2$ at $t=0$ to strongly resemble a black hole of total energy $\frac{E_1V}{V_1}$ in region $R$ but to also strongly resemble a black hole of total energy $\frac{E_2V}{V_2}$ in region $\bar R$.  This can be seen, for example, by considering the thermofield-double-like state defined by a Euclidean path integral where the period of Euclidean time is tuned independently in $R$ and $\bar R$ to obtain the desired energies and using a Euclidean version \cite{Marolf:2016lml} of the fluid-gravity correspondence \cite{Bhattacharyya:2008jc,Hubeny:2011hd,Rangamani:2009xk}.  The corresponding Renyi problem was recently discussed in \cite{Dong:2018seb}. }.

It is now manifest that we will find $S_1(\bar E_1) = \frac{\bar A_1}{4G}$ and $S_2(\bar E_2) = \frac{\bar A_2}{4G}$, though this can also be verified by direction computation using \eqref{eq:BTZbL1} and \eqref{eq:BTZbL2} in the limit of large $E = \frac{r_+^2}{8G}$.  Furthermore, the standard deviation $\tilde \sigma_-$ of our fixed-area discussion is easily extracted from the two-point functions \eqref{eq:sigma1}, \eqref{eq:sigma2}, and \eqref{eq:crosscor}.  At leading order in large $r_+$ we find
\begin{equation} \label{eq:heatConstant}
\frac{\tilde \sigma_-^2}{32 G} \approx \frac{(\pi^2-\eta^2)r_+}{8G\pi}.
\end{equation}

It thus remains only to compute \eqref{eq:Kdef} and compare with \eqref{eq:heatConstant}.  In terms of the parameter $\eta$ from \ref{Sec:BTZ} and the bulk Newton constant $G$ and the energy $E_1$, Cardy's formula \eqref{eq:Cardy} becomes
\begin{align}
S(\bar{E_1}) &= \frac{\pi+\eta}{4G}\sqrt{\frac{2\pi}{\pi+\eta}E_{1}},  \ \ \ {\rm and} \\
S(\bar{E_2}) &= \frac{\pi-\eta}{4G}\sqrt{\frac{2\pi}{\pi-\eta}(E-E_{1})}.
\end{align}
Using \eqref{eq:Kdef} then gives
\begin{equation} \label{eq:heatConstant2}
K = \frac{(\pi^2-\eta^2)r_+}{8G\pi},
\end{equation}
which agrees with \eqref{eq:heatConstant} as desired. So in the relevant limit our analysis does indeed reproduce the results of \cite{Murthy:2019qvb}.

\subsection{Comparison with a simple quantum RT transition}

In the above sections we have examined corrections to the RT entropy near RT phase transitions.  However, such phase transitions are very similar to the phase transitions associated with quantum extremal surfaces discussed in e.g. \cite{Penington:2019npb, Almheiri:2019psf,Penington:2019kki,Almheiri:2019qdq}.  Let us in particular consider the simple model described in section 2 of \cite{Penington:2019kki}, which considers a black hole in Jackiw-Teitelboim gravity with an end-of-the-world brane behind the horizon.  The end-of-the-world-brane can appear in any of $k$ flavors.  There is then a {\it quantum} RT phase transition associated with whether the entropy $\ln k$ of the state on the end-of-the-world brane exceeds the Bekenstein-Hawking entropy $S_{BH}$ of the black hole.  When $S_{BH}$ is the larger of the two, the (quantum) RT surface is the emptyset and the entire spacetime lies in the entanglement wedge of the boundary.  In contrast, if the end-of-the-world brane entropy is larger, the quantum RT surface lies instead at the black hole horizon and an `island' \cite{Almheiri:2019hni} forms inside.

Although this is technically a quantum-RT transition, quantum mechanics plays very little role in the discussion.  In particular, for the non-trivial extremal surface the entropy is well approximated by $A/4G$.  And for the trivial extremal surface, the (generalized) entropy is effectively a constant determined by the choice of end-of-the-world brane state.  It may thus be reasonable to expect that our arguments above would apply to this case as well.  We confirm this below, though we leave a full discussion of quantum phase transitions for future work.

In particular, in the semiclassical limit of large temperature $1/(\beta G) \gg 1$ and with large end-of-the-world brane tension $\mu_{EOW} \gg 1/(\beta G)$ , the details of their phase transition are studied in appendix F of \cite{Penington:2019kki} via a careful computation using the replica trick.  At the phase transition, the actual entropy is again found to be smaller than $A/4G$ by a correction
\begin{equation}
\label{eq:Wcoast}
\Delta_{-1/2} S=\sqrt{\frac{2\pi}{\beta G}}.
\end{equation}
We wish to verify that this result also follows from \eqref{eq:entcor} if we simply set $A_2 = \ln k$ (without fluctuations).  As a result, $4 \sigma_-^2 = \sigma_1^2$ and it remains only to determine the width of fluctuations in the horizon area $A_1$.

This width can be extracted from their $n$-replica partition functions
\begin{equation}
Z_{n}=e^{S_{0}} \int d s \rho(s) y(s)^{n},
\end{equation}
where
\begin{equation}
\rho(s)=\frac{s}{2 \pi^{2}} \sinh (2 \pi s),
\end{equation}
\begin{equation}
y(s)=e^{-\frac{\beta G s^{2}}{2}} 2^{1-2 \mu_{EOW}}\left|\Gamma\left(\mu_{EOW}-\frac{1}{2}+i s\right)\right|^{2}.
\end{equation}

In the limit $\mu_{EOW}\gg 1/\beta G$, the integrand can be approximated by
\begin{equation}
\label{eq:integrand}
\rho(s) y(s)^{n} \sim \frac{s}{2 \pi^{2}} y(0)^{n} e^{2 \pi s-n \beta G s^{2} / 2}.
\end{equation}
The saddle point is
\begin{equation}
s^{(n)}=\frac{2 \pi}{n \beta G}.
\end{equation}
We may thus define an on-shell action $I_n$ by inserting  $s^{(n)}$ into the exponent of \eqref{eq:integrand} to find
\begin{equation}
I_n=\frac{2\pi^2}{n\beta G}.
\end{equation}
The $n$-replica saddle-points should represent smooth geometries, but taking a ${\mathbb Z}_n$ quotient of such geometries should give spacetimes with a single boundary and a ${\mathbb Z}_n$ conical defect.  The fixed-defect-angle action $I_1(n)$ in such cases is generally $I_n/n$ (see \cite{Lewkowycz:2013nqa} and also \cite{Dong:2019piw} for further details of such actions).  We thus find
\begin{equation}
\label{eq:quotientI}
I_1(n):=\frac{I_n}{n}=\frac{2\pi^2}{n^2 \beta G},
\end{equation}
where the conical defect tension $\mu$ satisfies
\begin{equation}
n=\frac{1}{1-4\mu G}.
\end{equation}

It now straightforward to analytically continue the result \eqref{eq:quotientI} to all real $\mu$. As in section \ref{Sec:Correction}, the  variance of the RT area $A_1$ can be obtained by taking the second derivative of $I_1$ with respect to $\mu$:
\begin{equation}
\label{eq:bif}
\sigma_1^2=\left(\frac{\partial^2 I_1}{\partial \mu^2} \right)_{T\rightarrow 0}=\frac{64\pi^2 G}{\beta }.
\end{equation}
Inserting \eqref{eq:bif} into \eqref{eq:entcor} gives \eqref{eq:Wcoast} in agreement with \cite{Penington:2019kki}

\section{Discussion}
\label{Sec:Disc}

Our work above studied corrections to the Ryu-Takayanagi entropy of holographic systems near an RT-phase transition in the semiclassical limit.  Using a decomposition into fixed-area states we found that, when a so-called diagonal approximation holds, the result can be written in the form \eqref{eq:entcor}.  In particular, at the phase transition where the mean value $\bar A_1 -\bar A_2$ vanishes, we find a correction of order $G^{-1/2}$ controlled by the width $\sigma_- = G^{1/2} \tilde \sigma_-$ of the fluctuations in $(A_1-A_2)/2$.  This correction is parametrically larger than corrections associated with the entropy of bulk quantum fields.

However, it also decays exponentially in $|A_1-A_2|$ as one moves away from the transition.  In particular, just as in \cite{Murthy:2019qvb}, with this correction the entanglement becomes a smooth function of all parameters.  The RT `phase transition' has thus become a crossover already at this level of analysis, though in the limit $G \rightarrow 0$ the crossover happens very quickly and one recovers the sharp transition of the standard classical RT-surfaces. 

This behavior is very different from the $O(N)$ corrections described in \cite{Donnelly:2019zde} for 2d Yang-Mills.  Although that theory admits a `bulk' closed string expansion, the strings are light.  As a result, they give rise to D-brane-like (and thus $O(N)$) contributions to general entropies \cite{Donnelly:2016jet,Donnelly:2018bef}, regardless of proximity to a phase transition.  In contrast, stringy modes appear to play no role in our effect.

The interesting question that we have not addressed is just when this diagonal approximation should hold. We conjecture that it holds for arbitrary holographic states, but this remains to be verified.  What we have done in this regard is to compare our \eqref{eq:entcor} with the exact results at this order that are known in two cases.  The first was the large mass limit of (pure microstates of) BTZ black holes.  If we take the black hole to be in an energy eigenstate, then since the dual theory is conformal this limit is equivalent to the large volume limit studied by Murthy and Srednicki in \cite{Murthy:2019qvb}.  We found in section \ref{Sec:Matching} that our results coincide with theirs in the desired limit.

Now, one might ask if the condition that the black hole is an energy eigenstate might enforce our diagonal approximation even if the approximation were to fail more generally.  And indeed, for classical saddles that contribute to holographic Renyi computations, one expects the areas $A_1,A_2$ to be functions of the energies $E_1,E_2$ of the two parts of the system ($R$ and $\bar R$).  As a result, since $E_1 + E_2 =  E$ is fixed, given two pairs of areas, ($A_1,A_2$) and ($A_1', A_2'$) either the pairs coincide ($A_1=A_1'$ and $A_2=A_2'$), or both areas differ ($A_1 \neq A_1'$ {\it and also} $A_2 \neq A_2'$).  But as described in section \ref{Sec:Diag}, the saddles that give possible off-diagonal contributions require at least one area in each Renyi copy to coincide with one area in the next.  So there are no off-diagonal contributions with $A_1 \neq A_1'$ {\it and also} $A_2 \neq A_2'$ and the diagonal approximation should hold.

On the other hand, one can give a state-counting argument that generalizes the argument of \cite{Murthy:2019qvb} to generic states with a given expectation value of the energy, but which leaves the result unchanged\footnote{We thank Chaitanya Murthy and Mark Srednicki for sharing their notes on this point.}.  This removes the above constraint and allows off-diagonal saddles to contribute.  Yet we continue to find agreement with the computations of section \ref{Sec:BTZ}.  Indeed, our analysis made no use of any assumption regarding the width of fluctuations in the total energy of the black hole.

We take this as encouraging evidence in favor of our conjecture.  However, one can expect the diagonal approximation to fail for carefully chosen non-generic states, and there remains the possibility that at least some holographic states are non-generic in just the required way -- though this cannot be the case for pure microstates of BTZ.

We also performed what appears to be an independent check on our conjecture by comparing \eqref{eq:entcor} with the results of \cite{Penington:2019kki} for their quantum RT-transition.  While we have not analyzed quantum transitions in detail, one would expect analogous results to hold, and especially so for the special case considered in \cite{Penington:2019kki} where the quantum contributions are fixed and do not fluctuate.  And indeed we find our \eqref{eq:entcor} to exactly reproduce the $G^{-1/2}$ correction of \cite{Penington:2019kki}.

A by-product of the computations in our examples was to investigate the cutoff dependence of fluctuations in RT-areas.  In AdS$_3$, we found RT-surfaces anchored to the boundary to have fluctuations whose variance is of order $-\ln \delta$, and thus whose width is of order $\sqrt{-\ln \delta}$.  They thus diverge as $\delta \rightarrow 0$, but do so more slowly than the RT-lengths themselves (which are of order $\ln \delta$).   Furthermore, given two extremal surfaces $\gamma_1,\gamma_2$ anchored at the same boundary points, the {\it difference} in their lengths $L_1-L_2$ has finite (cutoff-independent) fluctuations as $\delta \rightarrow 0$.

It is straightforward to see that similar results must hold in complete generality and in all dimensions.
First, recall from section \ref{Sec:Examples} that fluctuations are related to expected RT-areas via
\begin{equation}
\label{eq:fluctdiff}
\langle A_i A_j \rangle_{0,0} - \langle A_i\rangle_{0,0} \langle A_j \rangle_{0,0} = - \frac{\partial}{\partial \mu_i}  \langle A_j \rangle_{\mu_1,\mu_2}\Big |_{\mu_1 = \mu_2=0}.
\end{equation}
In general, the divergences (or cutoff-dependences) of the variance RT-area fluctuations will agree with those of RT-areas $A$ at general tensions $\mu$, so that the width of such fluctuations scales like $A^{1/2}$.  This result is also to be expected physically, as the fluctuations should be local.  Since uncorrelated fluctuations add in quadrature, summing such fluctuations over all area elements of the RT-surface must again give fluctuations in the total area $A$ that scale like $A^{1/2}$.

In contrast, the cancellation of divergences that occurs in fluctuations of $A_1-A_2$ occurs precisely because the surfaces $\gamma_1,\gamma_2$ largely coincide near the AdS boundary, so that correlations between their area-fluctuations are naturally strong. That fluctuations of  $A_1-A_2$ will always be finite can be seen by recalling that any two extremal surfaces $\gamma_1,\gamma_2$ with the same boundary anchor set $\partial R$ in fact coincide near the boundary to all orders in the Fefferman-Graham expansion that give divergent contributions to $A_1$ and $A_2$ \cite{Taylor:2016aoi}.  Since this is the case for all smooth geometries with arbitrary matter sources, it will remain true in the conical limit where the sources become cosmic branes.  Thus $A_1-A_2$ is manifestly finite at general tensions $\mu_1,\mu_2$. Using \eqref{eq:fluctdiff} to write
\begin{equation}
\label{eq:fluctdiff}
\langle (A_1 - A_2)^2 \rangle_{0,0} - \langle A_1-A_2 \rangle_{0,0}^2  = \left( \frac{\partial}{\partial \mu_2} - \frac{\partial}{\partial \mu_1}\right)  \langle A_1 - A_2 \rangle_{\mu_1,\mu_2}\Big |_{\mu_1 = \mu_2=0},
\end{equation}
we see immediately that the desired fluctuations are finite as well.  The same argument indicates that one should be able to construct a holographically-renormalized bulk action for spacetimes with finite-tension cosmic branes anchored on the boundary, and similarly for spacetimes with boundary-anchored fixed-area surfaces.  We hope to return to the explicit construction of such actions in subsequent work.

It would also be interesting to explore other properties of fluctuations about holographic bulk saddles.  In particular, we saw above that fluctuations smooth out the classically-sharp RT phase transition of the entanglement entropy into a smooth crossover.  But in addition to this entropy, RT phase transitions also control the size and shape of the bulk entanglement wedge that can be recovered from a given boundary region $R$.  Fluctuations in bulk geometry should thus play a key role in smoothing out such transitions in the bulk reconstruction map.  Indeed, a natural extrapolation of our use of the diagonal approximation in section \ref{Sec:Correction} would be to {\it also} assume that we may approximate  the bulk reconstruction map at any $\bar A_1 - \bar A_2$
by using $\rho_D$ from \eqref{eq:rhoD}, and taking the map  to be the standard one determined by $\min (A_1,A_2)$ for each term in the sum over fixed-areas $A_1,A_2$.  This seems like to follow from the diagonal conjecture for entropy via a suitable generalization of the arguments in \cite{Jafferis:2015del} and \cite{Dong:2016eik}, though we leave exploration of the implications this conjecture and full justification for future work.

\paragraph{Acknowledgments}
It is a pleasure to thank Xi Dong for many discussions related to this material.  We also thank Chris Akers, Henry Maxfield, Chaitanya Murthy, Geoffrey Penington, Pratik Rath, Mark Srednicki, Douglas Stanford, and Xiaoliang Qi for useful discussions. This material is based upon work supported by the Air Force Office of Scientific Research under award number FA9550-19-1-0360.  It was also supported in part by funds from the University of California.  Finally, DM thanks the Kavli Institute for Theoretical Physics for their hospitality during a portion of this work.  As a result,
this research was also supported in part by the National Science Foundation under Grant No. NSF PHY-1748958.

\appendix
\section{Action calculations for one interval case}
\label{app}
This appendix derives the Euclidean action \eqref{eq:oneint} for the one-interval case. The action contains three parts: the Einstein-Hilbert term, the Gibbons-Hawking term and the counterterms. The action also depends on the choice of cutoff $\delta$ introduced in section \ref{Sec:SingleInterval}.

While one could calculate the bulk action using the metric (\ref{Poincare2}), it turns out to be easier to use the cylindrical coordinates in which the metric takes the form
\begin{equation}
\label{eq:cyl}
ds^2=(1+r^2)dx^2+\frac{dr^2}{1+r^2}+\alpha^2 r^2d\phi^2
\end{equation}
Our cutoff spacetime is then bounded by the extremal surfaces $x=-\frac{1}{2}\alpha L_0$ and $x=\frac{1}{2}\alpha L_0$, which in the Poincar\'e ball coordinates are anchored to the boundary cutoff surfaces described in section \ref{Sec:SingleInterval}.  In order to arrive at a description where the coordinate ranges are independent of $\alpha$, we introduce $\tilde x = x/\alpha \in [-L_0/2,L_0/2]$ which yields
\begin{equation}
\label{eq:cyl2}
ds^2=\frac{dr^2}{1+r^2}+\alpha^2\left[(1+r^2)d\tilde{x}^2+ r^2d\phi^2\right].
\end{equation}

The metric \eqref{eq:cyl2} can be written in the Fefferman-Graham form by defining
\begin{equation}
z:=\frac{2}{\alpha}\frac{1}{r+\sqrt{1+r^2}},
\end{equation}
which yields
\begin{equation}
\label{FG}
ds^2=\frac{1}{z^2}\left(dz^2+(1+\alpha^2 z^2/4)^2 d\tilde x^2+(1-\alpha^2 z^2/4)^2d\phi^2\right).
\end{equation}
Here $z=0$ is the AdS boundary and $z=2/\alpha$ is the $\phi$-axis.
The associated boundary metric is just a cylinder of length $L_0$ and circumference $2\pi$.  For convenience, we may now identify the extremal surfaces $\tilde x = \pm L_0/2$ so that the boundary becomes a torus. While actions $I$ computed in this conformal frame may differ from those computed in the round conformal frame, the difference arises only from the conformal anomaly.  Since the anomaly is the same for each state (i.e., for each $\alpha$), this contributes only an overall normalization constant (which might depend on $\delta$ and $\lambda$) to our probabilities $P(\alpha) \sim e^{-I}$, and in any case the normalization must be later fixed to yield $\int d\alpha P(\alpha) =1$.

We are thus free to use the above toroidal frame for any value of $\lambda$. The action consists of an Einstein-Hilbert term (with a cosmological constant), a Gibbons-Hawking term, and a counter-term.  Since $R-2\Lambda=-4+16\pi\mu G\delta(x^\mu-x^\mu_{\text{string}})$, the Einstein-Hilbert term may be further divided into two parts.
The contribution from string itself is clearly
\begin{equation}
I_{\text{string}}= -\mu \alpha L_0= \frac{(\alpha-1)\alpha L_0}{4G}.
\end{equation}
Since a radial cutoff at $z=\epsilon$ yields $r= \frac{1}{\alpha \epsilon}(1 - \alpha^2 \epsilon^2/4 +O(\epsilon^4))$, the Einstein-Hilbert (with cosmological constant) contribution from the region away from the string is
\begin{equation}
\begin{aligned}
I_{EH1}&=-\frac{1}{16\pi G}\int d^3 x \sqrt{g} (-4)\\
&=    \frac{\alpha L_0}{2G}  \int^{r(\epsilon)}_{0} \alpha r dr =  \frac{\alpha^2 L_0}{4 G} \left(\frac{1}{\alpha^2 \epsilon^2} - \frac{1}{2} \right).
\end{aligned}
\end{equation}
To calculate the Gibbons-Hawking term, we first need to calculate the extrinsic curvature on the surface $r=r(\epsilon)$. The unit normal to that surface is
\begin{equation}
n^\mu \partial_\mu=  \sqrt{1 + r^2} \partial_r,
\end{equation}
so the trace of the extrinsic curvature is
\begin{equation}
\begin{aligned}
K&=n^\rho \partial_\rho \ln \sqrt{g}+\partial_\rho n^\rho\\
&= \sqrt{1 + r^2} \partial_r \ln (\alpha r) + \partial_r \sqrt{1 + r^2} = 2 + O(r^{-4})
= 2 + O(\epsilon^{4})
\end{aligned}
\end{equation}
Since a constant $r$ surface has area $2\pi \alpha^2 L_0 r\sqrt{1 + r^2} = 2\pi \alpha^2 L_0 r^2\sqrt{1 + r^{-2}} = \frac{2\pi}{\epsilon^2}  L_0 + O(\epsilon^2)$,
the Gibbons-Hawking term is
\begin{equation}
\begin{aligned}
I_{GH}&=-\frac{1}{8\pi G}\int d^2x\sqrt{h}K\\
&=-\frac{ L_0}{2G\epsilon^2}  + O(\epsilon^{2}),
\end{aligned}
\end{equation}
where $\sqrt{h}$ is the area element of the induced metric on the surface $r=constant$.
Finally, the counterterm is
\begin{equation}
\begin{aligned}
I_{CT}&=\frac{1}{8\pi G }\int d^2x \sqrt{h} \\
&=
\frac{L_0}{4 G\epsilon^2}  + O(\epsilon^2).
\end{aligned}
\end{equation}
Summing these terms and taking $\epsilon \rightarrow 0$ gives the total action \eqref{eq:oneint}.

\bibliographystyle{jhep}
	\cleardoublepage

\renewcommand*{\bibname}{References}

\bibliography{FixedArea}

\end{document}